\documentclass[12pt]{article}
\usepackage{amsmath}
\usepackage{graphicx}
\usepackage{enumerate}
\usepackage{cauchy}
\usepackage{natbib}
\usepackage{url} % not crucial - just used below for the URL 

\usepackage{color}
\usepackage{amssymb}
\usepackage{mathtools}
\usepackage{paralist}
\usepackage{comment}
\usepackage{array}
\usepackage{breqn}

\usepackage{algorithm}
\usepackage[algo2e, boxed, ruled, vlined]{algorithm2e}

\usepackage{xr}
\newcommand{\blind}{1}

\theoremstyle{plain}
\newtheorem{thm}{Theorem}[section]

\theorembodyfont{\rmfamily}
\newtheorem{rem}[thm]{Remark}
\newtheorem{cor}[thm]{Corollary}
\newtheorem{ass}{Assumption}
\newtheorem{prop}[thm]{Proposition}
\newcommand{\bq}{\begin{eqnarray*}}
\newcommand{\eq}{\end{eqnarray*}}

\newcommand{\ii}{\imath}

\newcommand{\verk}{\stackrel{{\cal D}}{\longrightarrow}}

\renewcommand{\baselinestretch}{1.4}

\renewcommand{\BBAA}{and}

\newcommand{\indep}{\,\raisebox{0.05em}{\rotatebox[origin=c]{90}{$\models$}}\,}
\newcommand{\n}{^{(n)}}
\newcommand{\bo}[1]{\boldsymbol{#1}}
\newcommand{\Cov}{\mathop{\mathsf{Cov}}\nolimits}
\usepackage{stackengine}
\stackMath
\newcommand\tenq[2][1]{%
\def\useanchorwidth{T}%
\ifnum#1>1%
\stackunder[0pt]{\tenq[\numexpr#1-1\relax]{#2}}{\scriptscriptstyle\thicksim}%
\else%
\stackunder[1pt]{#2}{\scriptscriptstyle\thicksim}%
\fi%
}

\date{}

\begin{document}

\def\spacingset#1{\renewcommand{\baselinestretch}%
{#1}\small\normalsize} \spacingset{1}

%%%%%%%%%%%%%%%%%%%%%%%%%%%%%%%%%%%%%%%%%%%%%%%%%%%%%%%%%%%%%%%%%%%%%%%%%%%%%%

\if1\blind
{
  \title{\bf Consistent Distribution--Free Affine--Invariant Tests  for the Validity \\ of Independent Component Models}
  \author{Marc Hallin\thanks{
    Marc Hallin gratefully acknowledges  the support of the Czech Science Foundation grants GA\v{C}R22036365 and GA24-10078S. }\hspace{.2cm}\\
    ECARES and Department of Mathematics, Université Libre de Bruxelles\\  and \\ 
    Czech Academy of Sciences, Prague, Czech Republic\\
     \\
    Simos G. Meintanis \\
    Department of Economics, National and Kapodistrian University
of Athens\\ and \\ 
Pure and Applied Analytics,  North--West University, Potchefstroom, South Africa\\ \vspace{-1mm}\\
    and \vspace{-1mm} \\ \\
    Klaus Nordhausen \\
    Department of Mathematics and Statistics,  University of Jyv\"askyl\"a}
  \maketitle
} \fi

\if0\blind
{
  \bigskip
  \bigskip
  \bigskip
  \begin{center}
    {\bf Consistent Distribution--Free Affine--Invariant Tests  for the Validity of  Independent Component Models}
\end{center}
  \medskip
} \fi

\bigskip
\begin{abstract}
We propose a family of tests 
of the validity of the assumptions underlying independent component analysis methods. 
The tests are formulated as L2--type procedures based on characteristic functions and involve weights;    a proper choice of these weights and the estimation method for the mixing matrix yields consistent and affine-invariant tests. Due to the complexity of the asymptotic null distribution of the resulting test statistics,  implementation is based on permutational and resampling strategies. This leads to distribution-free procedures regardless of whether these procedures are performed on the estimated independent components themselves or the componentwise ranks of their components. A Monte Carlo study involving various estimation methods for the mixing matrix, various weights, and a competing test based on distance covariance is conducted under the null hypothesis as well as under alternatives. A real-data application demonstrates the practical utility and effectiveness of the method.
\end{abstract}

\noindent%
\vfill

%\newpage
\spacingset{1.9} % DON'T change the spacing!
\section{Introduction}\label{sec_1}
\subsection{Testing the validity of the independent component model}

Consider the {\it independent component model} (ICM)
\begin{equation} \label{ICM}
{\boldsymbol {X}}={\boldsymbol{\mu}}+{\boldsymbol{\Omega}}{\boldsymbol {Z}} 
%{\boldsymbol {X}}_j={\boldsymbol{\mu}}+{\boldsymbol{\Omega}}{\boldsymbol {Z}}_j \quad j=1,...,n,
\end{equation}
whereby the $p$-dimensional random vector  ${\boldsymbol {X}}\coloneqq (X_1,...,X_p)^\top %\in \mathbb R^p
, \ p>1$ is linearly associated with a latent %centered
  random vector 
 $
{\boldsymbol {Z}}={\boldsymbol {Z}}(\boldsymbol{\mu}, \boldsymbol{\Omega})\coloneqq (Z_1(\boldsymbol{\mu}, \boldsymbol{\Omega}),...,Z_p(\boldsymbol{\mu}, \boldsymbol{\Omega}))^\top = (Z_1,...,Z_p)^\top %\in \mathbb R^p
$ 
 of centered independent components, i.e., ${\boldsymbol {Z}}(\boldsymbol{\mu}, \boldsymbol{\Omega})$ is enjoying the property of {\it total independence}:\footnote{While {\it total independence}  \eqref{indep} \vspace{-2mm} implies {\it pairwise independence}---namely,
 $Z_j\indep Z_k$ 
%Z_j (\boldsymbol{\mu}, \boldsymbol{\Omega})\indep Z_k(\boldsymbol{\mu}, \boldsymbol{\Omega})$ 
%\quad \forall 
for all $j\neq k=1,\ldots,p
$ 
 where$\indep$, as usual, denotes stochastic independence---the converse, of course, is not true. 
}  the $p$-dimensional  distribution  of ${\boldsymbol {Z}}(\boldsymbol{\mu}, \boldsymbol{\Omega})$ 
 is the product of the  (marginal) distributions of its components $Z_1(\boldsymbol{\mu}, \boldsymbol{\Omega}),\ldots, Z_p(\boldsymbol{\mu}, \boldsymbol{\Omega})$.  
 
  Denoting by $\varphi_{\ell}, \ \ell=1,...,p$ and~ $\varphi$ the characteristic functions (CFs) of $Z_\ell({\boldsymbol{\mu}},{\boldsymbol{\Omega}})$ and~$\boldsymbol Z({\boldsymbol{\mu}},{\boldsymbol{\Omega}})$, respectively,  the property of total independence is equivalent to 
 \begin{equation}\label{indep} 
\varphi(\boldsymbol t)=\prod_{\ell=1}^p \varphi_\ell(t_\ell),\qquad  %\forall
 \boldsymbol t=(t_1,...,t_p)^\top \in \mathbb R^p.   
\end{equation}
The linear structure of (\ref{ICM}) involves a mean vector ${\boldsymbol{\mu}} \in \mathbb R^p$ and a $(p\times p)$ matrix $\boldsymbol{\Omega}$, often termed the {\it mixing matrix}, which belongs to the space $\mathbb M^p$ of full-rank $(p\times p)$ matrices.

The independent component model (ICM) is underlying the broad class of methods known as {\it  independent component analysis} (ICA).  A widely used method---a Google search\footnote{google.com accessed on 05/03/2024} returns no less than  532,000,000 links!---ICA  originally was developed in the engineering literature on signal processing, where it has countless applications. It recently   also attracted much interest  in finance and economics:  see, for instance,
 \citet{%BackWeigand1997, 
 ComonJutten2010, GarciaFerrerGonzalezPrietoPena2012, MattesonTsay2017, GourierouxMonfortRenne2017, Hai2020}, or  \citet{MiettinenMatilainenNordhausenTaskinen2020}, 
 to quote only a few.   All these applications crucially rely on the validity of the ICM assumptions \eqref{ICM}--\eqref{indep} in % are satisfied for
  some observed sample~${\boldsymbol {X}}_1,\ldots,{\boldsymbol {X}}_n$ of i.i.d.\ copies of~$\boldsymbol X$. Somewhat surprisingly, though, no test of the validity of these assumptions, i.e.,  of the the null hypothesis\vspace{-1mm} 
\begin{equation}\label{null} {\cal{H}}_0\!:~\mbox{the random vector} \  \boldsymbol{X}   \ \mbox{ is such that \eqref{ICM} and \eqref{indep} hold     for  some $({\boldsymbol{\mu}},{\boldsymbol{\Omega}}) \in \mathbb R^p \times \mathbb M^p$}\vspace{-1mm}
\end{equation}
is available in the literature except for a procedure proposed in   \cite{MattesonTsay2017} who are using the notion of {\it distance covariance} (see Section~\ref{sec_6} for details on  the distance covariance approach) and resampling-based critical values. Their test statistic, however,  depends on the arbitrary ordering of the components of~$\boldsymbol Z$, which is not a desirable property.     Our objective  here is to develop  tests of ${\cal{H}}_0$  which, as we shall see, are universally consistent and outperform the Matteson and Tsay procedure. 

Our tests are based on CFs and the fact that ${\cal{H}}_0$ %  in \eqref{null} 
holds if and only if there exists ${\boldsymbol{\mu}}$ and~${\boldsymbol{\Omega}}$ such that \eqref{indep} is satisfied. More precisely, our test statistics are weighted measures of the discrepancy between the  (empirical) joint CF and the product of marginal ones---a classical and quite natural measure of total dependence that can be traced back, e.g., to  \cite{CsorgoHall1982} and  is also considered in \cite{ChenBickel2005} in an estimation context.    Nobody  ever used these measures of total dependence  as   test statistics nor provided their asymptotic distributions under the assumption of total independence. The reason for this is that  these asymptotic distributions   (the distributions of  infinite sums of chi-squared random variables weighted by the eigenvalues of a complicated integral operator) are hopelessly untractable: indeed, they  not only depend  
 on the actual distribution of the observations but also on the way (FOBI, JADE, FastICA, \ldots\,) 
the mixing matrix is estimated. They cannot be used in the construction of asymptotic critical values and hence are entirely useless in practice.  This is probably the reason why these very natural characteristic-function-based statistics  so far have not been used as  test statistics. 

The main and nontrivial theoretical novelty of this contribution  is to show that these impracticable asymptotic critical values can be dispensed with and replaced with exact permutational  ones. These permutational critical values, however,  follow from a non-standard permutational scheme---not the usual $n!$ permutations of the observations, which leave the test statistics invariant! Moreover, the validity of our  permutational approach is anything but straightforward, since it involves  permutations (in dimension $p$, there are $(n!)^{p-1}$ of them)   of the components of estimated residuals that are not i.i.d.\ under ${\cal{H}}_0$. 

Our tests, unlike the Matteson and Tsay ones,  are affine-invariant and have exact size~$\alpha$ uniformly over~${\cal H}_0$, irrespective of the actual  (absolutely continuous) distribution of the observations and the demixing method  (FOBI, JADE, FastICA, \ldots\,)  adopted for deriving the residuals $\widehat{\bf Z}$. They  involve weights $W$; when choosing them,  we pay special attention to their impact on the computation of the corresponding test statistics, a feature that is particularly important in case of high-dimensional ICMs. 
% Finally, our tests also  have exact size~$\alpha$ uniformly over~${\cal H}_0$, irrespective of the actual  (absolutely continuous) distribution of the observations, hence fully distribution-free and affine-invariant---despite the preliminary estimation of  ICM residuals $\widehat{\boldsymbol{Z}}_j$'s which are not i.i.d.\ under the null. 
%However, we show that,   provided that~$\widehat{\boldsymbol\mu}$ and $\widehat{\boldsymbol\Omega}$ are symmetric functions of the ${\boldsymbol X}_j$'s, the~$\widehat{\boldsymbol{Z}}_j$'s enjoy an exchangeability property that allows for permutational critical values. 

Now, beyond its ICM structure,  ${\cal{H}}_0$ does not specify anything about $\boldsymbol X$; in particular, the distribution of~$\boldsymbol Z$, hence that of $\boldsymbol X$,   remains completely unspecified under ${\cal{H}}_0$  as well as under the alternative (which includes    arbitrary   dependencies between the components  of~$\boldsymbol X$). In principle, that distribution even could be discrete,   although we are focu\-sing on the absolutely continuous case  for simplicity.  The problem, therefore, is of a semiparametric nature, where the nuisance parameters are unspecified distributions. Ranks, in that context, naturally come into the picture, and we also propose %a class of 
  rank-based tests of ${\cal{H}}_0$ which  for the same reason as above,   admit fully distribution-free exact critical~values.

%To the best of our knowledge, this problem of testing the validity of the ICM has only briefly been touched by \citet{MattesonTsay2017}, %Matteson and Tsay~(2017), 
%who are using the notion of distance covariance. Their test statistic, however,  depends on the arbitrary ordering of the components of~$\boldsymbol Z$, which is not a desirable property. We refer to Appendix~\ref{sec_6} for details on  this distance covariance approach.

While the problem of testing bivariate independence has been treated quite extensively in the literature, the results on testing total independence are less abundant; see, however, \citet{BlumKieferRosenblatt1961,Deheuvels1981,Csorgo1985,Kankainen1995}.  
The related problem of testing vector independence---independence between vectors of finite dimension---recently  attracted  renewed
interest: we refer to %the reader to% the articles by 
\citet{RoySarkarGhoshGoswami2020,GenestNeslehovaRemillardMurphy2019,HerwartzMaxand2020,
%BilodeauNangue2017,
ShiDrtonHan2022}  
for reviews of the existing literature. CF-based   methods in this context date back to \citet{CsorgoHall1982,Csorgo1985,Feuerverger1993,KankainenUshakov1998}; 
for more recent contributions, see \citet{MeintanisIliopoulos2008,FanDeMicheauxPenevSalopek2017,PfisterBuhlmannScholkopfPeters2017,ChakrabortyZhang2019}, 
among others. In all these tests, an empirical contrast between the joint and the product of marginal CFs is considered. The advantages, in this context,  of the CF-based approach over competing methods   based, e.g.,  on  distribution function are confirmed by the success of distance covariance methods and their relation \citep{Sejd13} to RKHS statistics; see \citet{FP19}, \citet{SR23}, and \citet{ChenMeintanisZhu2019} for reviews on distance covariance and related approaches, and~\citet{ErikssonKoivunen2003} and~\citet{ChenBickel2005} for a discussion of their advantages in the ICM context.

As for the advantages of rank-based procedures in the presence of 
%already mentioned, the underlying distributions, under  ${\cal{H}}_0$ as well as under the~alterna\-tive,  remain largely 
unspecified distributions, we refer to \citet{ShiHallinDrtonHan2022}, who show that pseudo-Gaussian methods such as Wilks' classical test for vector independence \citep{Wilks1935},  although asymptotically valid,   severely over-reject under skewed distributions.  %and for moderately large dimensions. 
Rank-based independence testing methods therefore are a natural solution and have been thoroughly investigated in the bivariate context---see, e.g., Chapters II.4.11 and III.6 of \citet{hajekSidak1967}. With the recent introduction of measure transportation-based concepts of multivariate ranks and signs \citep{ChernozhukovGalichonHallinHenry2017,HallinBarrioCuestaAlbertosMatran2021}, 
 this %rank-based
  approach to bivariate independence testing has been extended to arbitrary dimensions   \citep{ShiDrtonHan2022,ShiHallinDrtonHan2022,ShiHallinDrtonHan2023,GhosalSen2022}.   
 The problem we are facing here, again, is trickier. Independence, in %all 
 these references, is to be tested between observable quantities  which under the null are i.i.d.\ so that the distribution-freeness of rank-based statistics is straightforward while we are dealing with unobservable 
 variables~$Z_{j\ell}$, the values of which have to be estimated as~$\widehat{Z}_{j\ell}$. These ``estimated residuals''  are no longer i.i.d.\ under the null, and handling them requires additional~care.

%In the ICM context,  CF-based 
%estimation procedures  have been developed by \citet{ErikssonKoivunen2003}  and \citet{ChenBickel2005} but  the  corresponding tests   have not been touched. The present work aims to fill this gap by investigating the properties of CF-based tests for the ICM. 
 
Still in the ICM  context,  rank-based tests of vector independence have been proposed by \citet{OjaPaindaveineTaskinen2016}. %Oja et al.~(2016). 
Their problem is quite different from ours, though: instead of the validity of the ICM assumptions, these authors are testing the independence of two given subvectors of~$\boldsymbol X$ under maintained   ICM assumptions.

%STOPPED HERE

\subsection{The test statistics}

Let ${\boldsymbol {X}}_1,\ldots,{\boldsymbol {X}}_n$, with  ${\boldsymbol {X}}_j\coloneqq (X_{j1},\ldots,X_{jp})^\top$, $j=1,...,n$  denote a sample of $n$ independent copies of $\boldsymbol X$. To implement  the test,   since we only observe ${\boldsymbol {X}}_j$  
and not ${\boldsymbol {Z}}_j$,
the latent variables $({\boldsymbol {Z}}_j, \ j=1,...,n)$ first need to be estimated from the data.    Specifically, our test statistic is a function of 
 the ``estimated ICM residuals" 
\begin{equation}\label{res}
\widehat{{\boldsymbol {Z}}}_j=\widehat{{\boldsymbol {Z}}}_j(\widehat{\boldsymbol{\mu}}_n,\widehat{\boldsymbol{\Omega}}_n)\coloneqq  \widehat{{\boldsymbol{\Omega}}}^{-1}_n \left({\boldsymbol {X}}_j-\widehat {{\boldsymbol{\mu}}}_n\right)\eqqcolon (\widehat Z_{j1},...,\widehat Z_{jp})^\top, \ j=1,...,n
\end{equation} 
obtained by plugging estimators $\widehat{\boldsymbol{\mu}}_n$ and $\widehat{\boldsymbol{\Omega}}_n$   of ${\boldsymbol{\mu}}$ and ${\boldsymbol{\Omega}}$ into (\ref{ICM}).
%where $(\widehat{\boldsymbol{\mu}}_n,\widehat{\boldsymbol{\Omega}}_n)$ denote estimators of $({\boldsymbol{\mu}},{\boldsymbol{\Omega}})$.
It is well known, however, that the matrix  ${\boldsymbol{\Omega}}$ in (\ref{ICM}) is not uniquely identified, and identification constraints need to be imposed which, without any loss of generality,  actually define the parameter space for~${\boldsymbol{\Omega}}$. These  constraints being closely related to the construction and statistical properties of $\widehat{{\boldsymbol{\Omega}}}_n$, their discussion is postponed to Section~\ref{sec_2},  which is dealing with the estimation of  ${\boldsymbol{\Omega}}$.

%\end{document}

We now rapidly describe the proposed test statistics. Denote by 
\begin{equation} \label{joint} \varphi^{(n)}({\boldsymbol{t}})\coloneqq \frac{1}{n}
\sum_{j=1}^n {\rm{e}}^{\imath {\boldsymbol{t}}^\top{\widehat{\boldsymbol{Z}}}_j},\quad\boldsymbol{t}\in \mathbb R^p  \vspace{-2mm}
%, \ (\ii=\sqrt{-1}),
\end{equation} 
and  \vspace{-2mm}
\begin{equation} \label{marg} 
 \varphi^{(n)}_{\ell}(t_\ell)\coloneqq \frac{1}{n}
\sum_{j=1}^n {\rm{e}}^{\imath t_\ell \widehat Z_{j\ell}},\quad t_\ell\in \mathbb R ,\ \ \ell=1,\ldots, p\end{equation} 
(where $\imath$ stands for the imaginary root of $-1$) the joint and marginal empirical CFs, respectively,  of the estimated  ICM resi\-du\-als~$\widehat {\boldsymbol{Z}}_j$, %\coloneqq  {\boldsymbol{Z}}_j(\widehat{\boldsymbol{\mu}}_n, \widehat{\boldsymbol{\Omega}}_n)$,
$j=1,\ldots,n$.
%where 
%$\widehat{\boldsymbol{\mu}}_n$ and  $\widehat{\boldsymbol{\Omega}}_n$ are adequate  estimators of ${\boldsymbol{\mu}}$ and ${\boldsymbol{\Omega}}$ and

Simi\-larly, letting~${\bf R}\n_j\coloneqq (R\n_{j1},\ldots ,R\n_{jp})^\top$ 
where $R\n_{j\ell}$, $j=1,\ldots,n$,   stands for the rank of~$\widehat Z_{j\ell}$ among~$\widehat Z_{1\ell},\ldots, \widehat Z_{n \ell}$, $\ell =1,\ldots,p$, define the joint and marginal rank-based statistics 
\begin{equation} \label{jointR}
 \tenq{\varphi}^{(n)}_{{\bf J}}({\boldsymbol{t}})\coloneqq \frac{1}{n}
\sum_{j=1}^n {\rm{e}}^{\imath {\boldsymbol{t}}^\top{{\bf J}({\bf R}\n_j/(n+1))}},\quad\boldsymbol{t}\in\mathbb R^p  \vspace{-2mm}
%, \ (\ii=\sqrt{-1}),
\end{equation} 
 and  \vspace{-2mm}
\begin{equation} \label{margR}  \tenq{\varphi}^{(n)}_{{\bf J},\ell}(t_\ell)\coloneqq \frac{1}{n}
\sum_{j=1}^n {\rm{e}}^{\imath t_\ell J_\ell(  R\n_{j\ell}/(n+1))}, \quad t_\ell\in\mathbb R ,\ \ \ell=1,\ldots, p\end{equation} 
where  
$
{\bf J}({\bf R}\n_j/(n+1))\coloneqq  (J_1(R\n_{j1}/(n+1)),\ldots,J_p(R\n_{jp}/(n+1)))^\top
$ and  
 ${\bf J}\coloneqq  (J_1,\ldots,J_p)^\top $ denotes a $p$-tuple of {\it score functions} $J_\ell : (0,1)\to \mathbb R$. 
Clearly, \eqref{jointR} and \eqref{margR} are the joint and marginal empirical CFs, respectively,  of the {\it scored ranks}~$J_\ell(R\n_{j\ell}/(n+1))$ of the estimated  ICM resi\-du\-als~$\widehat {\boldsymbol{Z}}_j$, $j=1,\ldots,n$.  

Our tests are rejecting   the null hypothesis ${\cal{H}}_0$  in \eqref{null} for large values of the test statistics 
\begin{equation}\label{TS} T_{n,W}\coloneqq  n \int_{\mathbb R^p} |D_n({\boldsymbol{t}})|^2   \: W(\boldsymbol t) {\rm{d}}{\boldsymbol{t}}
\quad\text{and}\quad  \tenq{T}_{n,{\bf J},W}\coloneqq  n \int_{\mathbb R^p} |\tenq{D}_{n,{\bf J}}({\boldsymbol{t}})|^2   \: W(\boldsymbol t) {\rm{d}}{\boldsymbol{t}},
\end{equation}  
%\begin{equation}\label{TS1} S_{n}\coloneqq  \sqrt{n} \sup_{\boldsymbol t \in \mathbb R^p} |D_n({\boldsymbol{t}})|,\quad \text{ and }\quad\tenq{S}_{n,{\bf J}}\coloneqq  \sqrt{n} \sup_{\boldsymbol t \in \mathbb R^p} |\tenq{D}_{n,{\bf J}}({\boldsymbol{t}})|\end{equation} 
(of the Cram\' er-von Mises type) where 
\begin{equation}\label{D} D_n(\boldsymbol t)\coloneqq \varphi^{(n)}({\boldsymbol{t}})-\prod_{\ell=1}^p
\varphi^{(n)}_{\ell}(t_\ell)\quad \text{ and }\quad
\tenq{D}_{n,{\bf J}}(\boldsymbol t)\coloneqq \tenq{\varphi}^{(n)}_{{\bf J}}({\boldsymbol{t}})-\prod_{\ell=1}^p
\tenq{\varphi}^{(n)}_{{\bf J},\ell}(t_\ell)\end{equation}
%In \eqref{TS}--\eqref{D},  the joint characteristic function of the vector $\boldsymbol Z=(Z_1,...,Z_p)^\top$ is estimated by 
%\begin{equation} \label{joint} \varphi_n({\boldsymbol{t}})=\frac{1}{n}
%\sum_{j=1}^n e^{\ii {\boldsymbol{t}}^\top{\widehat{\boldsymbol{Z}}}_j}, %\ (\ii=\sqrt{-1}),\end{equation} 
%based on the observations $\widehat {\boldsymbol{Z}}_j=(\widehat Z_{j1},...,\widehat Z_{jp})^\top$, $j=1,2,\ldots,n$, on $\boldsymbol Z$, while the marginal characteristic function of the component  $Z_\ell$ is estimated by 
%\begin{equation} \label{marg}  \varphi_{n\ell}(t_\ell)=\frac{1}{n}
%\sum_{j=1}^n e^{\ii t_\ell \widehat Z_{j\ell}},\end{equation} using the observations $(\widehat Z_{1\ell},\ldots,\widehat Z_{n\ell})$ on $Z_\ell, \ \ell=1,...,p$.
with $\boldsymbol t\coloneqq (t_1,t_2,\ldots,t_p)^\top \in \mathbb R^p$ and  some weight function $W:{\boldsymbol t}\mapsto W({\boldsymbol{t}})$. Test statistics~of the Kolmogorov-Smirnov type    
 $S_{n}\coloneqq  \sqrt{n} \sup_{\boldsymbol t \in \mathbb R^p} |D_n({\boldsymbol{t}})|$
% \quad\text{ and }\quad
 and $\tenq{S}_{n,{\bf J}}\coloneqq  \sqrt{n} \sup_{\boldsymbol t \in \mathbb R^p} |\tenq{D}_{n,{\bf J}}({\boldsymbol{t}})|$ 
 could also be considered; see for instance \citet{Csorgo1985}. However, computing a  $\sup_{\boldsymbol t\in \mathbb R^p}$~runs into technical difficulties  (already apparent in \citet{Csorgo1985})   
 since the empirical CF converges weakly  
  only over compact neighborhoods of $\mathbb R^p$ and this supremum has to be taken over some discrete grid. These drawbacks appear to have a direct impact on finite--sample powers, and Kolmogorov--Smirnov type tests typically are less powerful than their Cram\'er--von Mises counterparts (%see 
\citet{HenzeHlavkaMeintanis2014}). We, therefore, only consider the~latter.

\subsection{Outline of the paper}
The rest of the paper unfolds as follows. In Section \ref{sec_2},  we consider estimation methods for the ICM parameters $\boldsymbol\mu$ and $\boldsymbol\Omega$. Section \ref{sec_3}   concentrates on particular conditions under which the test statistics are affine-invariant while at the same time can be expressed in convenient closed forms, and illustrates certain connections with test statistics based on distances between densities. 
In Section~\ref{sec_4}, we consider permutation and bootstrap resampling methods for the computation of critical values.  
The consistency of our tests is established in Section~\ref{sec_5}, while in Section~\ref{sec_6} we consider an alternative method of testing based on distance covariance.  Section~\ref{sec_7} reports the results of a Monte Carlo study on the finite-sample behavior of the new tests, with various implementation choices, e.g.,  different estimators, weight functions, scores, and permutational or bootstrap-based critical values. Section \ref{sec_8} is devoted to an empirical biomedical application. Section \ref{sec_9} concludes. The appendix contains some technical arguments (Appendix~\ref{AppA}), extra simulation results (Appendix~\ref{AppB}), and additional material on the real-data application of Section \ref{sec_8}  (Appendix~\ref{AppC}).

\section{Estimation of $\boldsymbol{\mu}$ and $\boldsymbol{\Omega}$}\label{sec_2}
\subsection{Estimation of $\boldsymbol{\mu}$: shift invariance}\label{sec21}
%Let us start with the test statistics 
An important feature of the test statistics \eqref{TS} is their invariance with respect to  shifts acting on $\boldsymbol{X}$. Let us show, in particular, that $\widehat{\boldsymbol\mu}_n$ does not influence the test statistics~\eqref{TS}. Let $D_n^0(\boldsymbol t)\coloneqq n^{-1}\sum_{j=1}^n  {\rm{e}}^{\imath  \boldsymbol t^\top \widehat{\boldsymbol \Omega}_n^{-1}{\boldsymbol X}_{j}}- \prod_{\ell=1}^p  n^{-1}\sum_{j=1}^n  {\rm{e}}^{\imath t_\ell (\widehat{\boldsymbol \Omega}_n^{-1}{\boldsymbol X}_{j})_\ell}$: clearly,~$D_n^0(\boldsymbol t)$ is obtained by letting $\widehat{\boldsymbol{\mu}}_n={\boldsymbol 0}$ in $D_n(\boldsymbol t)$. Then, 
\begin{eqnarray*}
  D_{n} (\boldsymbol t)  & = & 
\frac{1}{n}\sum_{j=1}^n {\rm{e}}^{\imath  \boldsymbol t^\top \widehat{\boldsymbol Z}_{j}}- \prod_{\ell=1}^p  \frac{1}{n}\sum_{j=1}^n {\rm{e}}^{\imath t_\ell \widehat Z_{j\ell}} \\ 
%\end{eqnarray} 
%\end{document}
%\\
 & = &
%e^{-\imath  \boldsymbol t^\top \widehat{\boldsymbol\Omega}_{n}^{-1}\widehat{\boldsymbol\mu}_n}
\frac{1}{n}\sum_{j=1}^n  {\rm{e}}^{\imath  \boldsymbol t^\top\widehat{\boldsymbol\Omega}_{n}^{-1}\left({\boldsymbol X}_j -\widehat{\boldsymbol\mu}_n \right)}
-\prod_{\ell=1}^p \frac{1}{n} \sum_{j=1}^n  {\rm{e}}^{\imath   t_\ell \left( \widehat{\boldsymbol\Omega}_{n}^{-1}\left({\boldsymbol X}_j -\widehat{\boldsymbol\mu}_n \right)\right)_\ell}
 \\
& = &   {\rm{e}}^{-\imath  \boldsymbol t^\top \widehat{\boldsymbol\Omega}_{n}^{-1}\widehat{\boldsymbol\mu}_n }
\frac{1}{n} \sum_{j=1}^n  {\rm{e}}^{\imath  \boldsymbol t^\top 
\widehat{\boldsymbol\Omega}_{n}^{-1}{\boldsymbol X}_j
}
- \prod_{\ell=1}^p   {\rm{e}}^{-\ii     t_\ell 
\left(\widehat{\boldsymbol \Omega}_{n}^{-1}\widehat{\boldsymbol\mu}_n\right)_\ell}
  \prod_{\ell=1}^p \frac{1}{n} \sum_{j=1}^n  {\rm{e}}^{\ii t_\ell  
  \left( \widehat{\boldsymbol\Omega}_{n}^{-1} {\boldsymbol X}_j\right)_\ell}
\\ 
& = &   {\rm{e}}^{-\imath  \boldsymbol t^\top \widehat{\boldsymbol\Omega}_{n}^{-1}\widehat{\boldsymbol\mu}_n }
\left(
\frac{1}{n} \sum_{j=1}^n  {\rm{e}}^{\imath  \boldsymbol t^\top 
\widehat{\boldsymbol\Omega}_{n}^{-1}{\boldsymbol X}_j
}
-
\prod_{\ell=1}^p \frac{1}{n} \sum_{j=1}^n  {\rm{e}}^{\ii t_\ell   \left( \widehat{\boldsymbol\Omega}_{n}^{-1} {\boldsymbol X}_j\right)_\ell}
\right)
%\\ 
%& = &
=   {\rm{e}}^{-\imath  \boldsymbol t^\top \widehat{\boldsymbol\Omega}_{n}^{-1}\widehat{\boldsymbol\mu}_n }  D_{n}^0 (\boldsymbol t),
\end{eqnarray*} 
where $\boldsymbol t =(t_1,\ldots,t_p)^\top$. Hence,
$%\[
\left\vert 
D_{n} (\boldsymbol t)
\right\vert
=
\left\vert
 {\rm{e}}^{-\imath  \boldsymbol t^\top \widehat{\boldsymbol\Omega}_{n}^{-1}\widehat{\boldsymbol\mu}_n }\right\vert \, \left\vert
D_{n}^0 (\boldsymbol t)
\right\vert
=
\left\vert
D_{n}^0 (\boldsymbol t)
\right\vert
$ %\]
and the test statistic $T_{n,W}$, which only depends on $\left\vert D_{n}(\cdot)\right\vert$, safely can be computed from  $D_{n}^0(\cdot)$ instead of $D_{n}(\cdot)$. This allows us to skip the estimation of $\boldsymbol\mu$.

The ranks of the ICM residuals $\widehat{{\boldsymbol {Z}}}_j $ being insensitive to location shifts, the same conclusion directly holds for the rank-based statistics $\tenq{T}_{n,{\bf J},W}$.

\subsection{Estimation of the unmixing matrix $\boldsymbol{\Omega}^{-1}$}\label{OmegaSec} 
 The ICA literature  proposes a   number of   estimators for the {\it unmixing matrix} $\boldsymbol \Omega^{-1}$ (see, e.g., \citet{ComonJutten2010,NordhausenOja2018}). Below, we are focusing on three of the most popular of them: FOBI (fourth order blind identification, \citep{Cardoso1989}), JADE (joint diagonalization of eigenmatrices, \citep{CardosoSouloumiac1993}), and symmetric FastICA \citep{HyvarinenOja1997}.   The main reasons for these choices are that they are computationally fast and widely used.

Before estimating  $\boldsymbol{\Omega}^{-1}$, however, one first needs  to impose identification constraints on~$\boldsymbol{\Omega}$. Let $\boldsymbol P$, $\boldsymbol J$, and $\boldsymbol D$  denote an arbitrary $p$-dimensional permutation matrix,  an arbitrary  $p$-dimensional sign-change matrix (a diagonal matrix with $\pm 1$ on its diagonal), and an arbitrary  $p$-dimensional scaling matrix (a diagonal matrix with strictly positive diagonal elements), respectively.
 Then, 
\[
\boldsymbol X = \boldsymbol \mu + \boldsymbol \Omega \boldsymbol Z =  \boldsymbol \mu + (\bo \Omega \bo P \bo J \bo D) (\bo D^{-1} \bo J \bo P^\top \boldsymbol Z) = \boldsymbol \mu + \boldsymbol \Omega^* \boldsymbol Z^*,
\]
which means that the order of the components, their signs, and their scales are not well identified. 
Except for these indeterminacies,  it can be shown,  however,  that $\bo Z$ is identifiable if at most one of its components is Gaussian. If $\bo Z$ has more than one Gaussian component, then the non-Gaussian ones are identifiable and the Gaussian ones are identifiable up to an additional rotation, see e.g.
\citet{TheisKawanabeMuller2011} for details. 
The scales of the independent components $Z_\ell$ are usually fixed by imposing~$\Cov(\bo Z) =~\!\bo I_p$ (the $(p\times p)$ identity matrix) and their order is determined by the specific ICA procedure used.  

Let $\bo X^{st} \coloneqq  \Cov(\bo X)^{-1/2}(\bo X - \bo \mu)$ denote the standardized (whitened) observed vector. Almost all ICA methods make use of the fact that $%\[
\bo X^{st} = \bo U^\top \bo Z
$ %\]
 for some $p \times p$ orthogonal matrix $\bo U$. The unmixing matrix then has the form $\bo \Omega^{-1} = \bo U^\top \Cov(\bo X)^{-1/2}$ \citep[see e.g.][]{MiettineNordhausenOjaTaskinen2014} and the various  ICA methods differ in the way the orthogonal matrix $\bo U$ is obtained.

Denote by $\overline{\bo X}$ and $\widehat \Cov$  the sample mean and the sample covariance matrix, respectively.  Then $\bo X_j^{st} \coloneqq  \widehat \Cov^{-1/2}(\bo X_j -\overline{\bo X})$, $j=1,\ldots,n$, and  FOBI estimate the orthogonal matrix~$\bo U$ as the $p\times p$ matrix consisting of the eigenvectors of the matrix of sample fourth moments~$
\widehat\Cov_4 \coloneqq  \frac{1}{p+2} \frac{1}{n}\sum_{j=1}^n  {\bo X_j^{st}}^\top  \bo X_j^{st} \bo X_j^{st} {\bo X_j^{st}}^\top.
$  
The FOBI unmixing matrix is well defined if all independent components have distinct kurtosis values, and the  components are usually ordered by decreasing  kurtosis  order.

To avoid the assumption of distinct kurtosis values, JADE computes  the $p^2$ cumulant matrices 
$%\[
\bo C^{kl} \coloneqq  \frac{1}{n} \sum_{j=1}^n  {\bo X_j^{st}}^\top \bo E^{kl} \bo X_j^{st} \bo X_j^{st} {\bo X_j^{st}}^\top - \bo E^{kl} - \bo E^{lk} - \delta_{kl} \bo I_p$, $1 \leq k \leq l \leq p
$ %\]
where $\bo E^{kl}$ is the $p \times p$ matrix with $(k,l)$th element  
  $\delta_{kl}$ (the usual Kronecker delta). The orthogonal matrix $\bo U$ of JADE  then is the common diagonalizer of   the %cumulant
   matrices~$\bo C^{kl}$, $1 \leq k \leq l \leq p$. JADE is well defined provided  that one component at most has kurtosis zero.

FOBI and JADE are called {\it algebraic} ICA approaches as they exploit properties of 
  cumulant matrices. An alternative popular approach consists of estimating  the rows\linebreak  of~$\bo U = \left(\bo u_1,\ldots,\bo u_p \right)$ by maximizing  some componentwise criteria of non-Gaussianity in a projection pursuit framework. The most widespread family of estimators of this type is  FastICA, where one has to choose the optimization criterion and  a measure  $G$ of non-Gaussi\-anity, and decide whether the rows of $\bo U$ are determined sequentially or simultaneously. 

The so-called {\it symmetric} FastICA determines the rows $\bo u_1,\ldots,\bo u_p$,  simultaneously and maximizes
 $%\[
\sum_{i=1}^p \mathsf E(|G(\bo u_i^\top \bo X^{st})|)
$ %\]
under the constraint $\bo U \bo U^\top = \bo I_p$ where the measure of non-Gaussianity $G$ can be any twice continuously
differentiable and nonquadratic function~$G$  satisfying $\mathsf E(G(Y)) = 0$ for $Y \sim N(0,1)$. Popular choices for $G$ are %for example 
\[
%\text
\mathsf{pow3:} \ G(x) \coloneqq (x^4 - 3)/4\quad\text{ and }\quad \mathsf{tanh:} \ G(x) \coloneqq  \log(\cosh(x)) - c_t, 
\]
where~$c_t \coloneqq  \mathsf E[\log(\cosh(Y))] \approx 0.375$ for $Y \sim N(0,1)$ is a normalizing constant (the names ($\mathsf{pow3}$ and $\mathsf{tanh}$) of these functions are   related to their derivatives, which are needed in the fixed-point algorithms used  for  estimation). For computational details, other variants, and other $G$ functions, see, for example, \citep{MiettinenNordhausenOjaTaskinenVirta2017,MiettinenNordhausenTaskinen2018}. 
%Miettinen at al (2017,2018).\\

Whether symmetric FastICA is consistent depends on the choice of the
function~$G$ involved.  It is known
\citep{MiettinenTaskinenNordhausenOja2015} that  $\mathsf{pow3}$  is
consistent provided that~there is at most one component with kurtosis
zero while, for most other $G$ functions,  consis\-tency conditions are
difficult to establish. For example, FastICA  with   $G=\ \mathsf{tanh}$~may fail for some densities \citep{Wei2014,VirtaNordhausen2017}.
However, it is  usually argued that these cases are so artificial that
$\mathsf{tanh}$, in practice, remains a good choice---actually, the 
most popular one in FastICA 
\citep{Hyvarinen1999,MiettinenNordhausenOjaTaskinenVirta2017}.

FOBI, JADE, and symmetric FastICA all 
are affine-equivariant in the sense that, under affine transformations of the data, the independent components at most change their signs and their order---which, however, can be fixed via adequate conventions. FOBI and JADE are consistent and symmetric FastICA (see the previous paragraph) is considered consistent.  When consistent,  all these estimators have limiting normal distributions, see \citet{MiettinenTaskinenNordhausenOja2015,MiettinenNordhausenOjaTaskinenVirta2017} for details. 

Clearly, FOBI, JADE, and symmetric FastICA   all require  moment assumptions.   ICA techniques  bypassing  the existence of finite moments do exist, though: see, e.g.,  Nordhausen, Oja, and Ollila (2008), Ilmonen and Paindaveine (2011), as well as Hallin and Mehta (2015). 
Unfortunately, they also are computationally much more intensive, which makes them impractical  in  permutational and bootstrapping approaches. 

\section{Weight functions}\label{sec_3}
In this section, we investigate  the computational issues related to our testing proce\-dures. In particular, we discuss the choice of the weight function $W$   in \eqref{TS} in connection with affine-invariance and 
tests based on contrasts between density estimators.   
%\textcolor{red}{This can be postponed after Section 3}
\subsection{Computational issues and affine-invariance} \label{invar}
The weight function    in \eqref{TS}  in principle, may be any integrable mapping $W$ from~$\mathbb R^p$ to the non--negative half--line.   As we shall see,  however, adequate choices of $W$ considerably simplify the computation of $T_{n,W}$ and $\tenq{T}_{n,{\bf J},W}$.  % of the test statistic $T_{n,W}$ in (\ref{TS}).

Denoting by  $w_1, \ldots, w_p$, a collection of $p$ symmetric (i.e., such that $w_\ell (t)=~\!w_\ell (-t)$) probability densities over $\mathbb R$,    set %as a weight function  
\begin{equation}\label{weight} W:\  {\boldsymbol t}=(t_1,\ldots,t_p)^\top \mapsto 
W(\boldsymbol t)\coloneqq \prod_{\ell=1}^p w_\ell(t_\ell).\end{equation} 
  This particular class of weight functions is in line with \citet{KankainenUshakov1998} and \citet{MeintanisIliopoulos2008}. 
Denoting by  
$%\[
{\cal{C}}_\ell(t)\coloneqq\int_{-\infty}^{\infty} \cos(t x) w_\ell(x) {\rm{d}}x$ %\]
%is the characteristic function corresponding to $W(\cdot)$, and   $\|\cdot\|$ denotes the usual Euclidean norm; see Fang et al.~(1990). 
 the CF of $w_\ell$ (since $w_\ell$ is symmetric, the imaginary part is zero), note that ${\cal{C}}_\ell(-t) = {\cal{C}}_\ell(t)$ for all~$t$. Then,~procee\-ding   as in \citet{KankainenUshakov1998} or \citet{MeintanisIliopoulos2008}, % Kankainen and Ushakov~(1998) or  Meintanis and Iliopoulos~(2008), 
we obtain %from~\eqref{TS}
\begin{equation} \label{TS11}
T_{n,W}= \frac{1}{n} \sum_{j,k=1}^n  \prod_{\ell=1}^p {\cal {C}}_\ell( {\widehat Z}_{jk,\ell})+\frac{1}{n^{2p-1}} \prod_{\ell=1}^p \sum_{j,k=1}^n {\cal{C}}_\ell( \widehat Z_{jk,\ell})-\frac{2}{n^p} \sum_{j=1}^n \prod_{\ell=1}^p \sum_{k=1}^n {\cal{C}}_\ell( \widehat Z_{jk,\ell})
\end{equation}
where  $\widehat Z_{jk,\ell}\coloneqq{\widehat Z}_{j\ell} - {\widehat Z}_{k\ell}\coloneqq
%$,  stands for the $\ell^{\rm{th}}$ component  of 
\left(\boldsymbol {\widehat Z}_{j}-\boldsymbol {\widehat Z}_{k}\right)_{\ell}$,\  $j,k=1,...,n$,  $\ell=1,..,p$.

This specific choice for $w_\ell$ considerably simplifies computations since there exist symmetric densities with particularly simple CFs, such as   the  stable densities, with~${\cal{C}}(t)={\rm{e}}^{-\gamma t^\eta}$, $\eta\in(0,2],\, \gamma>0$  \citep{Nolan2013}, or the  generalized Laplace densities,  with~${\cal {C}}(t)=(1+\gamma t^2)^{-\eta}$, $\eta,\gamma>0$ \citep{KozubowskiPodgorskiRychlik2013}. %, % Nolan (2013), and Kozubowski et al.~(2013), 
%respectively. 
For   $\eta=2$ in the stable density we get the Gaussian   CF~${\cal{C}}(t)={\rm{e}}^{-\gamma t^2}$, 
%\item if $w_\ell$ is a Laplace density, then  ${\cal{C}}_\ell(t)=(1+\gamma t^2)^{-1}$
%\item if $w_\ell$ is a Kotz--type density with $\eta=2$, then ${\cal{C}}_\ell(t)=e^{-t^2}(1-2t^2)$.
%\end{itemize}
while   $\eta=1$ in the generalized Laplace density yields $
{\cal{C}}(t)=(1+\gamma t^2)^{-1}$, the CF  of the classical Laplace density. 

Moreover, the resulting tests are affine-invariant, meaning that   
\begin{equation} \label{AI} 
T_{n,W}(\bo A\bo X_1+\bo b,...,\bo A\bo X_n+\bo b)=T_{n,W}(\bo X_1,...,\bo X_n)
\end{equation}
%holds true 
for any $\bo b \in \mathbb R^p$ and any full-rank  $(p\times p)$ matrix $\bo A$. To see this, first recall that,  in light of shift invariance (Section~\ref{sec21}), we can set $\bo b=\bo 0$.  The affine equivariance of all ICA methods considered in Section \ref{OmegaSec} implies 
 $\widehat {\bo \Omega}_{n,A}\coloneqq  \widehat {\bo \Omega}_{n}(\bo A\bo X_1,...,\bo A\bo X_n)=\bo A \bo J \bo P\widehat {\bo \Omega}_n$ 
 for some permutation matrix $\bo P$ and some sign-change matrix $\bo J$.   
Therefore,~from~\eqref{res}, we~have~that 
\begin{equation*}
\begin{split}
\widehat {\bo   Z}^{\bo A \bo X}_j(\widehat {\bo \Omega}_{n,\bo A}) &=(\widehat {\bo \Omega}_{n,\bo A })^{-1} \bo A {\bo X}_j=(\bo A \bo J \bo P \widehat {\bo \Omega}_{n})^{-1} \bo A  {\bo X}_j\\ &=\widehat {\bo \Omega}_{n}^{-1}  {\bo P}^{-1}  {\bo J}^{-1} {\bo A}^{-1} \bo A {\bo X}_j= \widehat {\bo \Omega}_{n}^{-1}   {\bo P}^\top   {\bo J} {\bo X}_j%\&=
={\bo J} {\bo P}^{\top}\widehat {\bo Z}^{\bo X}_j(\widehat {\bo \Omega}_{n}),
\end{split}
\end{equation*}
where we use the fact that $\bo J$  and $\bo P$ are a sign matrix and a permutation matrix, respectively, so that $\bo {J}^{-1}=\bo J$  and  $\bo {P}^{-1}=\bo P^\top$.  Clearly then, for the class of weights   in \eqref{weight}, $\bo J$  and $\bo P$ have no impact on $T_{n,W}$, and  affine invariance readily follows. %}

\subsection{Connection with density contrasts}
%\textcolor{red}{Simos, the following subsections need some rewriting in the light of the changes}
In this section, we illustrate a   connection of the CF--based test with tests based on density estimators that   leads to a interesting interpretation of the weight function. To this end, recall the Parseval identity
\begin{equation}\label{parseval}
\int_{\mathbb R^p}   \Big|\varphi_1({\boldsymbol{t}})-\varphi_{2}(\boldsymbol t)\Big|^2 {\rm{d}}{\boldsymbol{t}}=(2\pi)^p  \int_{\mathbb R^p}   \Big[f_1({\boldsymbol{t}})-f_{2}(\boldsymbol t)\Big]^2 {\rm{d}}{\boldsymbol{t}},  \end{equation} 
whereby   L2--type distances between CFs $\varphi_m, \ m=1,2$,  translate to  distances between the corresponding densities $f_m, \ m=1,2$.   Also recall  the definition of  {\it adjoint pairs} of densities: a  (univariate) density~$w_1$ is called {\it adjoint} to the density~$w_2$ with CF~${\cal{C}}_2$\linebreak   if~${\cal{C}}_2(x)=w_1(x)/w_1(0)$ (this implies that ${\cal{C}}_2$ is real, hence that $w_2$  is  symmetric).    Conversely,   $w_2$ then is adjoint to $w_1$ and $(w_1, w_2)$ is  called an {\it adjoint pair}. Typical examples of adjoint densities are the Gaussian ones (which are  self--adjoint), and the pairs consisting  of   Laplace and Cauchy densities; for more examples, see  \citet{Rossberg1995}. % Rossberg (1995). 

In this context,  weight functions can be interpreted  as (real-valued) CFs  rather than densities. Let $W$ be of the form\footnote{Recall that, if ${\cal{C}}$ is the CF of some random variable $\xi$, then ${\cal{C}}^2$ is the CF of $\xi_1+\xi_2$ where $\xi_1$ and~$\xi_2$ are independent copies of $\xi$.}  $W=\prod_{\ell=1}^p {\cal{C}}^2_\ell$  where ${\cal{C}}_1 ,\ldots,{\cal{C}}_p$ are real-valued CFs. The  test statistic~(\ref{TS}) then can be written~as    
\begin{equation}\label{adj} 
T_{n,W}=n \int_{\mathbb R^p} \Big| \varphi^{(n)}({\boldsymbol{t}})\prod_{\ell=1}^p {\cal{C}}_\ell(t_\ell)-
\prod_{\ell=1}^p \varphi^{(n)}_{\ell}(t_\ell) {\cal{C}}_\ell(t_\ell) \Big|^2 {\rm{d}}{\boldsymbol{t}}.
\end{equation}  
Let $F_n$ and $F_{n\ell}$,  $\ell=1,...,p$ denote the joint and marginal  empirical distribution functions of~$\widehat {\boldsymbol Z}_1,\ldots,\widehat {\boldsymbol Z}_n$.  Then, writing $\cal F_\ell$, $\ell=1,\ldots,p$ and ${\cal{F}}$ for the distribution functions corresponding to ${\cal{C}}_\ell$ and  $\prod_{\ell=1}^p {\cal C}_\ell(t_\ell)$, respectively,   $\varphi^{(n)}({\boldsymbol{t}})\prod_{\ell=1}^p {\cal{C}}_\ell(t_\ell)$ and $\varphi^{(n)}_{\ell}(t_\ell){\cal{C}}_\ell(t_\ell)$ are  the CFs associated with the convolutions $F_n \ast {\mathcal {F}}$ and  $F_{n\ell} \ast {\cal F}_\ell$, $\ell=1,...,p$, respectively. Applying the Parseval identity \eqref{parseval} in \eqref{adj}  yields 
$%\[
T_{n,W}=n (2\pi)^p   \int_{\mathbb R^p}   \left[ f_n({\boldsymbol{t}})-\prod_{\ell=1}^p f_{n\ell}(t_\ell)\right]^2 {\rm{d}}{\boldsymbol{t}}$, % \] 
where~$f_n$ is the density corresponding to   $F_n \ast {\mathcal {F}}$, and   $f_{n\ell}$  the density corresponding to~$F_{n\ell} \ast {\cal {F}}_\ell$, $\ell=1,...,p$. Thus, the test statistic $T_{n,W}$  takes the form of an L2 distance between the  (random) densities $f_n$ and %the product 
$\prod_\ell f_{n\ell}$, where $\prod_\ell {\cal{C}}_\ell$ and $\cal C_\ell$   act as convolution operators. 
%
%
%the first is the joint ``empirical density" while the second is the product of  ``marginal empirical densities", both properly convoluted with densities corresponding to the weight functions.  

\section{Critical values and distribution-freeness}\label{sec_4}
The asymptotic null distribution of the test statistic $T_{n,W}$ is highly non--trivial and consequently  cannot be used for practical implementation of the  tests: see  \citet{Kankainen1995, KankainenUshakov1998, PfisterBuhlmannScholkopfPeters2017} for the case of specified (known) $\bf \Omega$,   and  Appendix~\ref{AppA} for the case of our composite hypothesis ${\cal {H}}_0$  under which~$\bf \Omega$ needs to be estimated. 
  For this reason, we recur to permutational and  resampling techniques in order to compute critical values and calibrate the testing procedures. 
  %For more information on the asymptotic null behavior of~$T_{n,W}$  the reader is referred to

Throughout this section, we make the assumption (satisfied by any reasonable estimator) that $\widehat{\boldsymbol{\Omega}}_n$ is a symmetric function of the $n$ observations---that is, $\widehat{\boldsymbol{\Omega}}_n$ is invariant under permutations of the ${\boldsymbol{X}}_i$'s; recall (Section~\ref{sec21})  that $\boldsymbol{\mu}$ can be assumed to be $\boldsymbol{0}$ without loss of generality, hence needs not be estimated.  
\subsection{Permutational critical values% for $T_{n,W}$ and $S_n$
}\label{permcritv}

Denote by $\widehat{Z}_{(j)k}$, $j=1,\ldots,n$, $k=1,\ldots, p$ the $j$th order statistic of the $k$th compo\-nents~$(\widehat{Z}_{k1},\ldots, \widehat{Z}_{kn})$ of the estimated ICM residuals   $(\widehat{\boldsymbol{Z}}_{1},\ldots, \widehat{\boldsymbol{Z}}_{n})$. 
% and let  $\widehat{\boldsymbol{Z}}_{(\cdot)k}\coloneqq ({\widehat{Z}}_{(1)k},\ldots, {\widehat{Z}}_{(n)k})$, $k=1,\ldots,p$.   
Define the   {\it order statistic}  of~$\widehat{\boldsymbol{Z}}^{(n)}\coloneqq (\widehat{\boldsymbol{Z}}_{1},\ldots, \widehat{\boldsymbol{Z}}_{n})$ as  the $n$-tuple~$\widehat{\boldsymbol{Z}}_{(\boldsymbol{\cdot})}^{(n)} \coloneqq (\widehat{\boldsymbol{Z}}_{(1)},\ldots, \widehat{\boldsymbol{Z}}_{(n)})$ of vectors of estimated ICM  residual  values reordered from smallest to largest values  of their first components, i.e., such that~ $ (\widehat{\boldsymbol{Z}}_{(j)})_1=  \widehat{Z}_{(j)1}$, $j=1,\ldots,n$. With the actual ICM residuals ${\boldsymbol{Z}}^{(n)}\coloneqq ({\boldsymbol{Z}}_{1},\ldots, {\boldsymbol{Z}}_{n})$ substituted for the estimated ones, the notation ${\boldsymbol{Z}}_{(\boldsymbol{\cdot})}^{(n)}$ will be used in an obvious  similar way; contrary to $\widehat{\boldsymbol{Z}}_{(\boldsymbol{\cdot})}^{(n)}$, however, ${\boldsymbol{Z}}_{(\boldsymbol{\cdot})}^{(n)}$  is not observable. Note that, since $\widehat{\bf Z}=\widehat{\boldsymbol\Omega}^{-1}{\boldsymbol\Omega}{\bf Z}$ where~$\widehat{\boldsymbol\Omega}$ is both $\widehat{\boldsymbol{Z}}_{(\boldsymbol{\cdot})}^{(n)}$- and~${\boldsymbol{Z}}_{(\boldsymbol{\cdot})}^{(n)}$-measurable, conditioning on~$\widehat{\boldsymbol{Z}}_{(\boldsymbol{\cdot})}^{(n)}$ is equivalent to conditioning on~${\boldsymbol{Z}}_{(\boldsymbol{\cdot})}^{(n)}$. 

Denoting by $\boldsymbol{\pi}=(\pi_1,\ldots,\pi_p)$ an arbitrary $p$-tuple of permutations of $\{1,\ldots,n\}$ and by~${\bf P}_{\pi_\ell}$, $\ell = 1,\ldots,p$ the  $n\times n$ permutation matrix corresponding with $\pi_\ell$ (the matrix~with  entries~$\big({\bf P}_{\pi_\ell}\big)_{ij}={\delta(i, \pi_\ell^{-1}(j))}$, with $\delta$  %denotes 
the Kronecker delta function),~let
\begin{equation}\label{Zpi}
\widehat{\boldsymbol{Z}}^{\boldsymbol\pi}\coloneqq 
\left(
\begin{array}{c}
(\widehat{\boldsymbol{Z}}_{(\boldsymbol{\cdot})}^{(n)})_{1\boldsymbol{\cdot} }\, {\bf P}_{\pi _1}\\ 
\vdots \\ 
(\widehat{\boldsymbol{Z}}_{(\boldsymbol{\cdot})}^{(n)})_{p\boldsymbol{\cdot}}\, {\bf P}_{\pi _p}
\end{array}
\right).
\end{equation}
The  $\ell$th row in matrix $\widehat{\boldsymbol{Z}}^{\boldsymbol\pi}$ results from performing the permutation $\pi_\ell$ on row $\ell$ of  the order statistic $\widehat{\boldsymbol{Z}}_{(\boldsymbol{\cdot})}^{(n)}$.  As $\boldsymbol{\pi}=(\pi_1,\ldots,\pi_p)$ ranges over the $(n!)^p$ possible $p$-tuples~of permutations of $\{1,\ldots,n\}$,  $\widehat{\boldsymbol{Z}}^{\boldsymbol\pi}$ takes $(n!)^p$ values.  Denoting by $T^{\boldsymbol\pi}_{n,W}$ the test~statistic computed from~$\widehat{\boldsymbol{Z}}^{\boldsymbol\pi}$,~$T^{\boldsymbol\pi}_{n,W}$ similarly takes $(n!)^p$ possible values.  These $(n!)^p$ values are not all distinct, though: in particular, since $T_{n,W}$ is invariant under column permutations (i.e., ${\boldsymbol\pi}$ of the form~${\boldsymbol\pi}=(\pi,\ldots,\pi)$), these $(n!)^p$ possible values, with probability one, all have the same multiplicity $n!$. We therefore  restrict the range of ${\boldsymbol\pi}$ to the collection $\boldsymbol\Pi$\linebreak  of~$p$-tuples that are not column permutations (not  of the form~${\boldsymbol\pi}=(\pi,\ldots,\pi)$); then, with probability one, $T^{\boldsymbol\pi}_{n,W}$ has~$(n!)^{p-1}$ distinct possible values. Let us show that these $(n!)^{p-1}$ values, under the null and conditional on the order statistic $\widehat{\boldsymbol{Z}}_{(\boldsymbol{\cdot})}^{(n)}$ (equivalently, conditional on~${\boldsymbol{Z}}_{(\boldsymbol{\cdot})}^{(n)}$), are equiprobable. 

From \eqref{Zpi}, we have%, for  ${\boldsymbol\pi}^*\coloneqq (\pi_1^*,\ldots,\pi_p^*)$ with ,  
\begin{equation}\label{Zpi*}
\widehat{\boldsymbol{Z}}^{\boldsymbol\pi}= 
\left(
\begin{array}{c}
(\widehat{\boldsymbol{\Omega}}^{-1}\boldsymbol{\Omega})_{1\boldsymbol{\cdot} }\,
({\boldsymbol{Z}}^{(n)}\, {\bf P}_{\pi _1})\\ 
\vdots \\ 
(\widehat{\boldsymbol{\Omega}}^{-1}\boldsymbol{\Omega})_{p\boldsymbol{\cdot} }\,
({\boldsymbol{Z}}^{(n)}\, {\bf P}_{\pi _p})\end{array}
\right)
= 
\left(
\begin{array}{c}
(\widehat{\boldsymbol{\Omega}}^{-1}\boldsymbol{\Omega})_{1\boldsymbol{\cdot} }\,
({\boldsymbol{Z}}_{(\boldsymbol{\cdot})}^{(n)}\, {\bf P}_{\pi _1^*})\\ 
\vdots \\ 
(\widehat{\boldsymbol{\Omega}}^{-1}\boldsymbol{\Omega})_{p\boldsymbol{\cdot} }\,
({\boldsymbol{Z}}_{(\boldsymbol{\cdot})}^{(n)}\, {\bf P}_{\pi _p^*})\end{array}
\right)
\end{equation}
with ${\boldsymbol\pi}^*\coloneqq (\pi_1^*,\ldots,\pi_p^*)$ where $\pi^*_\ell = \pi {\scriptstyle \circ}\pi_\ell$ and $\pi$ is such that ${\bf Z}^{(n)}= {\bf Z}^{(n)}_{(\boldsymbol{\cdot})}{\bf P}_\pi$. 
 When $\boldsymbol\pi$ ranges over $\boldsymbol\Pi$, ${\boldsymbol\pi}^*$ similarly ranges over $\boldsymbol\Pi$. Under the null hypothesis and conditional  on~${\boldsymbol{Z}}_{(\boldsymbol{\cdot})}^{(n)}$, $({\boldsymbol{Z}}_{(\boldsymbol{\cdot})}^{(n)}\, {\bf P}_{\pi _1^*},\ldots, {\boldsymbol{Z}}_{(\boldsymbol{\cdot})}^{(n)}\, {\bf P}_{\pi _p^*})$ then takes, as  $\boldsymbol\pi$ ranges over $\boldsymbol\Pi$,  $(n!)^{p-1}$ possible values which are (conditionally) equiprobable. Now, \eqref{Zpi*} establishes an a.s.\ bijection between these~$(n!)^{p-1}$ possible values and those of $\widehat{\boldsymbol{Z}}^{\boldsymbol\pi}$ which therefore, still under the null hypothesis and conditional  on~${\boldsymbol{Z}}_{(\boldsymbol{\cdot})}^{(n)}$  (equivalently, on~$\widehat{\boldsymbol{Z}}_{(\boldsymbol{\cdot})}^{(n)}$), are also equiprobable. The claim about the $(n!)^{p-1}$ possible values of~$T^{\boldsymbol\pi}_{n,W}$ follows.

% In view of total  independence and absolute continuity, under the null hypothesis and conditional on   ${\boldsymbol{Z}}_{(\boldsymbol{\cdot})}^{(n)}$, the~$p\times n$ matrices $(\widehat{\boldsymbol{Z}}_{1},\ldots, \widehat{\boldsymbol{Z}}_{n})$ is uniformly distributed over the $(n!)^p$ values of 
%
%
%Under the null hypothesis, due to total  independence and absolute continuity, the distribution of the  sample (the $p\times n$ matrix with columns ${\boldsymbol X}_1,\ldots,{\boldsymbol X}_n$)  conditionally on the $p$ marginal order statistics $X_{(\cdot)1},\ldots, X_{(\cdot)p}$ is, with probability one,  uniform  over the~$(n!)^p$ matrices resulting from the $(n!)^{p-1}$ possible combinations of marginal permutations of these $p$ order statistics. 
%
%
%
%Each of these $(n!)^p$ matrices yields   a value of $\widehat{\boldsymbol\Omega}_n$, hence a value of our test statistics. However, the $n!$ permutations of the columns of any of them yield the same value of~$\widehat{\boldsymbol\Omega}_n$, hence the same value of our test statistics. Under  ${\cal H}_0$,  thus, still with probability one   and conditionally on the $p$ marginal order statistics of the sample,   our test statistics are uniform over $(n!)^{p-1}$ distinct values. 

As a consequence, permutational critical values can be constructed as the quantile of order $(1-\alpha)$ of that uniform distribution. While the test statistic $T_{n,W}$ itself is not distribution-free (since the   marginal order statistics of the sample are not), the resulting permutation test has exact\footnote{Since the permutational distribution is discrete, a randomization, in principle, might be required to match the exact size $\alpha$.  This, however, is highly negligible for large $n$;   we do not implement it in practice,  and   no longer   mention it in the sequel. } level~$\alpha$ and therefore is strictly distribution-free. Note, however, that the permutations involved here are not the usual $n!$ permutations of the observations. 

Even for moderate values of $n$ and $p$, enumerating the $(n!)^{p-1}$ possible permutational values of our statistics, of course,  is unfeasible. But these $(n!)^{p-1}$ values can be sampled, and the quantile of order $(1-\alpha)$ of that sample  still can be used as an exact$^3$~$\alpha$-level critical value  (this, when the number of permutations is small,  may induce a small loss of power but does not affect the size and distribution-freeness of the  test). 

\subsection{Bootstrap  critical values}% \textcolor{red}{STOPPED HERE 10H30}

As an alternative to the permutational approach of the previous section, one can also recur to the bootstrap to obtain critical values. 
%An essential condition, however, is that the bootstrap samples are created under the null hypothesis, i.e., under the ICM assumptions.
 An algorithm  (Algorithm~\ref{alg::boot}) producing  bootstrap $p$--values for $T_{n,W}$ is described below, where  the ICM assumptions are enforced by bootstrapping the $p$ estimated  independent components independently from each other. 

\bigskip

\begin{algorithm}\medskip
	
	Set the number of bootstrap samples $M$,  select a weight function $W$ and an ICA method,  and consider centered observed data  $\boldsymbol X_j, \ j=1,\ldots,n$;\;\\
	-- apply the ICA method to obtain $\widehat{\boldsymbol{\Omega}}_n$ and  $\widehat{\boldsymbol Z}_j = \widehat{\boldsymbol{\Omega}}^{-1}_n \boldsymbol X_j$, $j=1,\dots,n$;\;\\ 
    -- compute the test statistic $T=T_{n,W}$ based on $\boldsymbol X_j, \ j=1,\ldots,n$: \\
	\For{$m \in \{ 1, \ldots , M \}$}{
	- obtain bootstrap samples 
    $\boldsymbol Z_j^m = (Z_{j1}^m,\ldots,Z_{jp}^m)^\top$,
    $j=1,\ldots,n$, where $Z_{ji}^m$ is sampled from the empirical distribution of $(\widehat Z_{1i},\ldots, \widehat Z_{ni})$, $i=1,\ldots,p$;\;\\
    - compute ${\widehat{\boldsymbol{X}}}_j^m = \widehat{\boldsymbol{\Omega}}_n 
    \boldsymbol Z_j^m$, $j=1,\dots,n$;\;\\
    - compute the test statistic $T^m=T_{n,W}$ based on 
    ${\widehat{\boldsymbol{X}}}^m_j,\  j=1,\ldots,n$;\;
	%\If{spatial\_resampling = TRUE}{Replace $\boldsymbol Z^{*k}$ by a full spatial bootstrap sample. See text for details.;}
	%Compute $\boldsymbol X^{*k} \leftarrow \hat{\boldsymbol \Gamma} \bo Z^{*k} $ and $t^k \leftarrow t_r(\boldsymbol X^{*k})$;\;
}
	-- return the $p$--value$:\coloneqq [\#\{T^m \geq T\} + 1]/(M + 1)$.\smallskip
	
	\caption{Bootstrapping algorithm for the $p$--value of the $T_{n,W}$-based test of the~ICM~assumptions. }
	\label{alg::boot}
\end{algorithm}

Simulations indicate that he resulting bootstrap critical values only slightly differ from the permutational ones. 
In the simulation section below, we provide guidelines for the choice of the number $M$ of bootstrap samples and permutations required for a good approximation of critical values; see also Appendix~\ref{AppB4}.

Critical values for the rank tests based on~$\tenq{T}_{n,{\bf J},W}$  
 follow along the same lines as   in Section~\ref{permcritv}, with the additional property that the $(n!)^{p-1}$ possible values of $\tenq{T}_{n,{\bf J},W}$ 
  no longer depend on the values of the marginal order statistics. The test statistics~$\tenq{T}_{n,{\bf J},W}$,   
  thus,  are conditionally  uniform over these non-random  $(n!)^{p-1}$ values---hence they are unconditionally uniform. It follows that  $\tenq{T}_{n,{\bf J},W}$,  
  unlike ${T}_{n,W}$,  
   is genuinely distribution-free under ${\cal H}_0$---and so are the resulting tests.

\subsection{The normal score test}
A classical choice for the score function~${\bf J}=(J_1,\ldots, J_p)$ is the normal or van der Waerden score $J_\ell= \Phi^{-1}$, $\ell = 1,\ldots,p$ where~$\Phi$, as usual, stands for the standard normal distribution function. Then, instead of~$\tenq{D}_{n,{\bf J}}(\boldsymbol t)$ as defined in \eqref{D}, 
one may like to consider 
 \[
 \tenq{D}_{n,{{\text{\rm vdW}}}}(\boldsymbol t)\coloneqq \tenq{\varphi}^{(n)}_{{\bf J}}({\boldsymbol{t}})-{\rm{e}}^{-\frac{1}{2}\Vert \boldsymbol{t}\Vert ^2}
 \quad\text{with}\quad 
 \Vert \boldsymbol{t}\Vert ^2\coloneqq \sum_{\ell =1}^p t_\ell ^2.
 \]
%(with~$\Vert \boldsymbol{t}\Vert ^2\coloneqq \sum_{\ell =1}^p t_\ell ^2$) by noting that 
%$\tenq{\varphi}^{(n)}_{{\bf J},\ell}(t_\ell)$, in this particular case, converges to the standard normal CF $t_\ell\mapsto {\rm{e}}^{-\frac{1}{2}t_\ell^2}$. 
%

\section{Consistency}\label{sec_5}

In this section, we establish the consistency of our tests under rather mild conditions on the weight function $W$; namely, we make the following assumption.
\begin{ass} \label{ass1} The weight function $W$ is such that
\begin{compactenum}
\item[(i)] $W({\boldsymbol{t}})>0$ Lebesgue--a.e.\ on $\mathbb{R}^p$;  \ \ (ii) $\int_{\mathbb R^p} W(\boldsymbol t) {\rm{d}}\boldsymbol t<\infty$.
%\item[(ii)]
\end{compactenum}
\end{ass}

The proof goes along   similar lines as in \citet{MeintanisIliopoulos2008}, 
with the   modifications required by  the estimation of the mixing matrix ${\boldsymbol{\Omega}}$. The estimator~$\widehat{\boldsymbol{\Omega}}_n$ is assumed to converge  strongly. 
More precisely,  we assume the following.    
\begin{ass} \label{ass2}
As $n\to\infty$, $\widehat{\boldsymbol{\Omega}}_n$  converges  strongly 
to some  ${\boldsymbol{\widetilde\Omega}}$, where ${\boldsymbol{\widetilde\Omega}}$ under the null hypothesis ${\cal{H}}_0$ coincides with the actual value ${\boldsymbol\Omega}_0$ of $\boldsymbol\Omega$ and, under the alternative under study, denotes some  {\it pseudo--true value}.\footnote{See, for instance, \citet{MattesonTsay2017}.} 
\end{ass}
Of course,   ${\boldsymbol{\Omega}}$,  ${\boldsymbol{\widetilde\Omega}}$, and  $\widehat{\boldsymbol{\Omega}}_n$ must satisfy the identification constraints, that is, it should belong to the    subset $\widetilde{\mathbb M}^p$ of matrices in $\mathbb M^p$ satisfying these constraints (see Section~\ref{OmegaSec}).  

Note that Assumption~\ref{ass2} is not just an assumption on $\widehat{\boldsymbol{\Omega}}_n$ but on $\widehat{\boldsymbol{\Omega}}_n$ and the particular  alternative under study. 
In the sequel, we denote by $\widehat{\boldsymbol{\Omega}}_n$ and~${\cal H}_1$, respectively, a sequence of estimators and the collection of alternatives under which $\widehat{\boldsymbol{\Omega}}_n$ satisfies  Assumption~\ref{ass2}. 

 Denote by $\widetilde \varphi$   the CF of~$\widetilde {\boldsymbol Z}=(\widetilde{Z}_1,\ldots,\widetilde{Z}_p)^\top\coloneqq\widetilde {\boldsymbol \Omega}^{-1} \boldsymbol X$ and by $\widetilde \varphi_\ell$   the CF of its~$\ell$th component 
$\widetilde Z_\ell$,  
 $\ell=1,...,p$. Then, 
 \begin{compactenum}
 \item[--] under %the null hypothesis
  ${\cal H}_0$, $\widetilde \varphi$ and $\widetilde \varphi_\ell$ coincide with $\varphi$ and $\varphi_\ell$, %and
   hence 
 $\widetilde \varphi ({\boldsymbol{t}})=\prod_{\ell =1}^p\widetilde \varphi_\ell(t_\ell)$ for all $\boldsymbol{t}\in\mathbb{R}^p$;
\item[--]  under the alternative
 ${\cal H}_1$, there exists some $\boldsymbol t_0=(t_{01},...,t_{0p})^\top \in \mathbb{R}^p$ %{\cal{B}}_0(T)$
such that 
\begin{equation}\label{H1}
  \widetilde \varphi(\boldsymbol t_0)\neq \prod_{\ell=1}^p \widetilde \varphi_\ell(t_{0\ell}).  \end{equation}  
\end{compactenum}

 \begin{prop} \label{cons}{\it 
Let Assumptions \ref{ass1} and  \ref{ass2} hold. Then, under any fixed alternative in~${\cal H}_1$, 
\begin{center}$T_{n,W} \rightarrow \infty$ {\rm a.s.}\ as $n\to \infty$.\end{center}
 }
 \end{prop}

\vspace{0.2cm}

{\bf{Proof}}. 
From (\ref{TS}), we have 
\begin{equation} \label{conve}
\frac{1}{n}{T_{n,W}}=\int_{\mathbb R^p} |D_n(\boldsymbol t)|^2 W(\boldsymbol t) {\rm{d}} \boldsymbol t\end{equation}
where $D_n(\cdot)$ is defined in (\ref{D}). 
Letting $\widehat{\boldsymbol{u}}_n\coloneqq  (\widehat{\boldsymbol{\Omega}}^{-1}_n)^\top\boldsymbol t$,    the joint empirical CF    (\ref{joint}) takes the form 
 $%$\[  
\varphi^{(n)}({\boldsymbol{t}})= \frac{1}{n}
\sum_{j=1}^n {\rm{e}}^{\imath \widehat{\boldsymbol u}^\top_n \boldsymbol X_j}\eqqcolon  \phi_n(\widehat{\boldsymbol{u}}_n)$, %\] 
where $\phi_n(\boldsymbol t)=
n^{-1}
\sum_{j=1}^n {\rm{e}}^{\imath \boldsymbol t^\top \boldsymbol X_j}$ is the empirical CF of    $\boldsymbol X_1,\ldots,\boldsymbol X_n$. By Assumption \ref{ass2}, as $n\to\infty$, %we have
$\widehat{\boldsymbol{u}}_n \longrightarrow ({\boldsymbol{\widetilde\Omega}}^{-1})^\top \boldsymbol t$~a.s., which,  invoking  the strong uniform consistency of the empirical CF over compact intervals (see \citet{Csorgo1981}, \citet{Csorgo1985}),  
implies that, pointwise in~$\boldsymbol t$, 
\begin{equation}\label{joint1}  \phi_n(\widehat{\boldsymbol{u}}_n) \longrightarrow   \mathsf E({\rm{e}}^{\imath \boldsymbol t^\top \widetilde {\boldsymbol \Omega}^{-1} \boldsymbol X})=
 \widetilde\varphi(\boldsymbol t)
, \ \mbox{a.s. as} \ n\rightarrow \infty.  
\end{equation}

Also, denoting by  $\widehat{\boldsymbol{\omega}}^{(-1)}_{n\ell}$   the $\ell^{\rm{th}}$ row, $\ell=1,...,p$, of  
$\widehat{\boldsymbol \Omega}^{-1}_n$,    write the marginal empirical CFs~(\ref{marg}) as  
$%\[   
 \varphi^{(n)}_{\ell}(t_\ell)= \frac{1}{n}
\sum_{j=1}^n {\rm{e}}^{\imath  \widehat{\boldsymbol u}^\top_{n\ell} \boldsymbol X_{j}}\eqqcolon   \phi_{n}(\widehat {\boldsymbol u}_{n\ell}), 
$ %\]
where $\widehat {\boldsymbol u}_{n\ell}\coloneqq  (\widehat{\boldsymbol \omega}^{(-1)}_{n\ell})^\top t_\ell$. 
Then, by Assumption 2, %we have~
$\widehat {\boldsymbol u}_{n\ell} \longrightarrow   (\widetilde {\boldsymbol \omega}^{(-1)}_{\ell})^\top t_\ell$ a.s., which, for the same  reasons as for  (\ref{joint1}), implies  
\begin{equation}\label{marg1}
\phi_{n}(\widehat {\boldsymbol u}_{n\ell}) \longrightarrow \mathsf E({\rm{e}}^{\imath t_\ell \widetilde {\boldsymbol \omega}^{(-1)}_\ell \boldsymbol X})= \widetilde \varphi_\ell(t_\ell), \ \mbox{a.s. as} \ n\to\infty,
\end{equation} 
where $(\widetilde {\boldsymbol \omega}^{(-1)}_{\ell}, \ \ell=1,...,p)$, stands for the $\ell^{\rm{th}}$ row  of    $\widetilde {\boldsymbol \Omega}^{-1}$.

It follows from (\ref{joint1}) and (\ref{marg1})  that \begin{equation} \label{Conv} D_n({\boldsymbol{t}})  \longrightarrow \widetilde \varphi(\boldsymbol t)  -\prod_{\ell=1}^p \widetilde \varphi_\ell(t_\ell)\eqqcolon \widetilde D(\boldsymbol t)
,  \ \mbox{a.s. as} \ n\rightarrow \infty.\end{equation} 
Taking (\ref{Conv}) into account, and since  $|D_n({\boldsymbol{t}})|^2\leq 4$,   an application  to  (\ref{conve})  of Lebesgue's  Dominated Convergence Theorem yields
%\begin{equation*} \label{conv2}
$ \frac{1}{n}{T_{n,W}}  \longrightarrow \int_{\mathbb R^p} |\widetilde D(\boldsymbol t)|^2 \: W(\boldsymbol t) {\rm{d}} \boldsymbol {t}\eqqcolon \Delta_W$  a.s.~as $n\to~\!\infty$. 
%\ \mbox{a.s. as} \ n \rightarrow \infty.\end{equation*}
In view of Assumption \ref{ass1}, $\Delta_W$  is strictly positive unless $\widetilde D(\boldsymbol t)=0$ $\boldsymbol{t}$-a.e., which 
holds if and only if~${\cal H}_0$ is true while, under the alternative, \eqref{H1} implies $\Delta_W >0$. The claim follows.\footnote{Actually, due to the continuity of CFs, \eqref{H1} holds over a neighborhood of ${\bo {t}}_0$.}\hfill$\square$

Turning to the rank-based tests, we have  very similar results. Denoting by $\widetilde{F}_\ell$  the distribution function of $\widetilde{Z}_\ell\coloneqq \big(\widetilde{\boldsymbol\Omega}^{-1}{\boldsymbol X}\big)_\ell$ under Assumption~\ref{ass2}, let  ${\cal H}_{1;{\bf J}}^T$ be the subset  of~${\cal H}_1$ under which there exists some $\boldsymbol t_0=(t_{01},...,t_{0p})^\top \in {\cal{B}}_0(T)$ 
such that 
\begin{equation}\label{Jind}
\widetilde \varphi_{\bf J}(\boldsymbol t_0)\neq \prod_{\ell=1}^p \widetilde \varphi_{{\bf J},\ell}(t_{0\ell})
\end{equation}
where  $
 \widetilde{{\varphi}}_{{\bf J}}$ and $\widetilde{{\varphi}}_{{\bf J},\ell}$ are the CFs of~$\big( J_1 (\widetilde{F}_1(\widetilde{Z}_1)),\ldots,J_p(\widetilde{F}_p(\widetilde{Z}_p))\big)^\top$ and~$J_\ell(\widetilde{F}_\ell(\widetilde{Z}_\ell))$, respectively. The $p$ variables $\widetilde{Z}_\ell$, $\ell =1,\ldots,p$ are {\it totally} independent iff the~$p$ variables~$\widetilde{F}_\ell(\widetilde{Z}_\ell)$  (which are uniform over $[0,1]$) are. If the scores are strictly monotone, thus, \eqref{Jind} holds for some $\boldsymbol t_0 \in {\mathbb R}^p$ under any alternative in ${\cal H}_1$. 
\begin{prop} \label{consR}{\it 
Assume
that the scores  
$J_\ell$, $\ell=1,\ldots,p$ are strictly  
monotone. 
Let Assumptions \ref{ass1} and  \ref{ass2} hold.  Then, under any fixed alternative in~${\cal H}_1$, 
$\tenq{T}_{n,{\bf J},W}\! \rightarrow~\!\!\infty$ {\rm a.s.}\ as~$n\to \infty$.
 }
 \end{prop}
 
 \noindent {\bf{Proof}}.  
 The proof heavily relies on the continuity of the characteristic function as a mapping from the space of distribution functions  (Theorem~3.6.1 in \citet{Lukacs1970}; Theorems~2.6.8 and~2.6.9 in \citet{Cuppens1975}; Theorem~1.6.6 in \citet{Bogachev2018}). To be precise, the  mapping from a distribution function to the corresponding CF defines a homeomorphism  between the space of distribution functions equipped with the topo\-logy~of weak convergence and the space of CFs equipped with the topology of pointwise convergence.
 
The rank-based statistics $
 \tenq{\varphi}^{(n)}_{{\bf J}}$ and $\tenq{\varphi}^{(n)}_{{\bf J},\ell}$ defined in \eqref{jointR} and~\eqref{margR} actually are the joint and marginal characteristic functions of the uniform distribution over the~$n$-tuple $$( J_1 (R\n_{j1}/(n+1)),  \ldots,J_p(R\n_{jp}/(n+1)))^\top, \quad j=1,\ldots,n.$$ 
Denote by ${\widetilde{F}}_\ell$   the distribution function of~$\widetilde{Z}_\ell\coloneqq \big(\widetilde{\boldsymbol\Omega}^{-1}{\boldsymbol X}\big)_\ell$.  Glivenko-Cantelli, combined with the conti\-nuous~mapping theorem and Assumption~\ref{ass2},  imply the %uniform
 weak convergence of this joint distribution function to the distribution function  of $\big( J_1 ({\widetilde{F}}_1({\widetilde{Z}}_1)),\ldots,J_p({\widetilde{F}}_p({\widetilde{Z}}_p))\big)^\top$. The~$\ell$th marginal distribution function  similarly  weakly converges to the distribution function of $J_\ell\big(\widetilde{F}_\ell\big(\widetilde{Z}_\ell)\big)$, $\ell = 1,\ldots,p$.  Hence, in view of the continuity  of CFs, $
 \tenq{\varphi}^{(n)}_{{\bf J}}$ and~$\tenq{\varphi}^{(n)}_{{\bf J},\ell}$ pointwise converge  to the CFs~$
 \widetilde{{\varphi}}_{{\bf J}}$ and~$\widetilde{{\varphi}}_{{\bf J},\ell}$ of~$\big( J_1 (\widetilde{F}_1(\widetilde{Z}_1)),\ldots,J_p(\widetilde{F}_p(\widetilde{Z}_p))\big)^\top$ and~$J_\ell(\widetilde{F}_\ell(\widetilde{Z}_\ell))$, respectively. For monotone score functions~$J_\ell$, \eqref{Jind} holds, under any alternative in ${\cal H}_1$, for some $\boldsymbol t_0\! \in\! {\mathbb R}^p$. The claim then follows along the same lines as in the proof of Proposition~\ref{cons}.~$\square$

An immediate  corollary to Propositions~\ref{cons} and \ref{consR} is the consistency of our~tests.

\begin{cor}{\it The tests based on $T_{n,W}$ and $\tenq{T}_{n,{\bf J},W}$,  
 under Assumptions~\ref{ass1} and~\ref{ass2} and  for monotone scores in $\tenq{T}_{n,{\bf J},W}$, 
are consistent as $n\to\infty$ against any fixed alternative in ${\cal H}_1$.  
}
\end{cor}

\section{A distance covariance test} \label{sec_6}
As mentioned in the Introduction, the problem of  testing the validity of the ICM model seldom has been considered in the literature. To the best of our knowledge, the only competitor to our tests is the  distance covariance-based method proposed by \citet{MattesonTsay2017}. 

 {\it Distance covariance},  a concept that goes back to \citet{SzekelyRizzoBakirov2007}, % Sz\'ekely et al.~(2007), 
is one of the most popular methods for testing independence between two random vectors. It is well known, see for instance \citet{ChenMeintanisZhu2019}, % Chen et al.~(2019), 
that distance covariance is a special case of the general formulation implicit in \eqref{TS} that suggests measuring independence by means of a weighted L2--type  distance between the estimated joint CF and the product of the corresponding marginal ones.  To see this, consider $n$   independent copies~$(\boldsymbol \xi_j ^\top,\boldsymbol \eta_j^\top)$, $j=1,...,n$,  
 of a pair~$(\boldsymbol \xi ^\top,\boldsymbol \eta^\top)$ of random vectors taking values in $\mathbb R^p$ and $\mathbb R^q$, respectively. The empirical distance covariance between~$\boldsymbol \xi$ and~$\boldsymbol \eta$ is  of the form 
\begin{equation}\label{DC} 
{\rm{dc}}_n(\boldsymbol \xi,\boldsymbol \eta) \coloneqq   \int_{\mathbb R^{p+q}} \big|\varphi^{(n)}_{\boldsymbol \xi,\boldsymbol \eta}({\boldsymbol{t}})-\varphi^{(n)}_{\boldsymbol \xi}(\boldsymbol t_p) \varphi^{(n)}_{\boldsymbol \eta}(\boldsymbol t_q)\big|^2   \: W(\boldsymbol t) {\rm{d}}{\boldsymbol{t}}\eqqcolon T_{n,{\rm dc}},\end{equation} 
with the weight function   $W(\boldsymbol t)\coloneqq c_{p,q}\|\boldsymbol t_p\|^{-(1+p)} \|\boldsymbol t_q\|^{-(1+q)}$, where $\|\cdot\|$ denotes the usual Euclidean norm and $c_{p,q}$ is an irrelevant constant; here $\varphi^{(n)}_{\boldsymbol \xi,\boldsymbol \eta}({\boldsymbol{t}})$ denotes the empirical joint CF of~$(\boldsymbol \xi ^\top_j,\boldsymbol \eta^\top_j)$, $j=1,\ldots,n$,  computed as in  \eqref{joint} at~$\boldsymbol t=(\boldsymbol t _p^\top,\boldsymbol t _q^\top)^\top$,  while
$$\varphi^{(n)}_{\boldsymbol\xi}(\boldsymbol t_p)\coloneqq \varphi^{(n)}_{\boldsymbol \xi,\boldsymbol \eta}((\boldsymbol t _p^\top,0,...,0)^\top)\quad\text{and}\quad  \varphi^{(n)}_{\boldsymbol \eta}(\boldsymbol t_q)\coloneqq \varphi^{(n)}_{\boldsymbol \xi,\boldsymbol \eta}((0,...,0,\boldsymbol t _q^\top)^\top)$$ 
denote the corresponding empirical marginal CFs.   This leads to the convenient expression  
\begin{equation}
{\rm{dc}}_n(\boldsymbol \xi,\boldsymbol \eta) %&=&
=\frac{1}{n^2} \sum_{j,k=1}^n \| \boldsymbol\xi_{jk}\|  \| \boldsymbol\eta_{jk}\|+\frac{1}{n^2}  \sum_{j,k=1}^n \|  \boldsymbol\xi_{jk}\|   \frac{1}{n^2} \sum_{j,k=1}^n \|  \boldsymbol\eta_{jk}\|
%\\ \nonumber &-&
-\frac{2}{n^3} \sum_{j=1}^n \sum_{k,\ell=1}^n   \|\boldsymbol\xi_{jk}\|  \|\boldsymbol\eta_{j\ell}\|,
\end{equation}
with $\boldsymbol \xi_{jk}\coloneqq\boldsymbol \xi_{j}-\boldsymbol \xi_{k}$ and $\boldsymbol \eta_{jk}\coloneqq\boldsymbol \eta_{j}-\boldsymbol \eta_{k}$, $j,k=1,...,n$.   Under the general form
~\eqref{DC}, the connection between ${\rm{dc}}_n$ and the test statistic $T_{n,W}$ in~\eqref{TS}   becomes quite evident.

The null distribution of ${\rm{dc}}_n$ depends on the distributions of $\boldsymbol\xi$ and $\boldsymbol\eta$, but distribu\-tion-free versions of distance covariance based on the so-called center-outward ranks and signs \citep{HallinBarrioCuestaAlbertosMatran2021} %(Hallin et al.~2021)  
have been proposed recently by \citep{ShiDrtonHan2022,ShiHallinDrtonHan2022}. % Shi et al.~(2021a and b)

 Distance covariance and its rank-based versions, however, are not of direct use in our context   since they are designed to measure dependence  between pairs of observed vectors only while we need tests for total independence. This can be taken care of by  adapting an idea suggested by \citep{MattesonTsay2017} %Matteson and Tsay (2017) 
 and defining 
\[
{\rm{DC}}_n\coloneqq n \sum_{\ell=1}^{p-1}  {\rm{dc}}_n(\boldsymbol {\widehat \zeta}_{\ell},\boldsymbol {\widehat \zeta}_{-\ell})
\eqqcolon T_{n,{\rm DC}},
\]
where $\boldsymbol {\widehat \zeta}_{\ell}$, $\ell=1,\ldots,p$ corresponds to the $\ell$th independent component and $\boldsymbol {\widehat \zeta}_{-\ell}$ contains the remaining $p-1$ components.

%\textcolor{magenta}{@Simos: are you sure? Shouldn'it be $\boldsymbol {\widehat \zeta}_{\ell}\coloneqq(\widehat Z_{1},...,\widehat Z_{\ell})^\top, \ \ell=1,...,p$ and $\boldsymbol {\widehat \zeta}_{\ell^+}\coloneqq (\boldsymbol {\widehat Z}_{\ell+1},...,\boldsymbol {\widehat Z}_{p})$? PLEASE CHECK} $\boldsymbol {\widehat \zeta}_{\ell}\coloneqq(\widehat Z_{1\ell},...,\widehat Z_{n\ell}), \ \ell=1,...,p$ and $\boldsymbol {\widehat \zeta}_{\ell^+}\coloneqq (\boldsymbol {\widehat \zeta}_{\ell+1},...,\boldsymbol {\widehat \zeta}_{p})$. \textcolor{red}{This is not even invariant under permutation of the $p$ components!}

Analogously to \eqref{jointR}-\eqref{margR}, a distribution-free version of  the distance--covariance test statistic ${\rm{DC}}_n$ can be obtained  %, at least asymptotically,
by computing it from the scored ranks of   the estimated  resi-\linebreak duals~$(\widehat Z_{j\ell}, \ \ell=1,...,p)$. 
And, similarly to our tests,  the distance--covariance test statistic is affine-invariant as soon as the ICA method used in estimation is affine-equivariant.

\section{Simulation results}\label{sec_7}
We conducted a simulation study to evaluate the performance of $T_{n,W}$ in a finite-sample setting to explore the following issues: (i) how does the test perform under ${\cal{H}}_0$ and how does the performance depend on the ICA method used to estimate the sources? (ii)~how does the performance depend on the score functions used, and is there a difference between using bootstrap- and permutation-based computation for the critical value? (iii) how do the tests compare to the competitor test of 
\citet{MattesonTsay2017} 
described in 
Section~\ref{sec_6}? (iv) how is the power with respect to various alternatives?

For that purpose, we consider the following three settings:
\begin{compactenum}
    \item[Setting 1:] a three-dimensional ICM with  uniform, exponential (with parameter $\lambda=~\!1$), and~$\chi^2_3$  components;    
    \item[Setting 2:] a three-dimensional spherical $t$-distribution with degrees of freedom (df) \linebreak
      in~$\{1, 2,  \ldots , 19, 20, 100, 500, \infty\}$, where  $t_\infty$ stands for  the Gaussian distribution (except for $t_\infty$, this setting does not yield  an ICM);
    \item[Setting 3:] a three-dimensional Clayton copula model with  tuning parameter  $\omega$  \linebreak in~$\{ 0, 0.1, \ldots, 1.4, 1.5\}$  (except for $\omega = 0$, this setting does not yield  an ICM).
\end{compactenum}

The ICA methods we used are the popular   FOBI, JADE, and FastICA methods based on~$\mathsf{tanh}$, which are all affine-equivariant, and thus due to affine--invariance of the test (see Section \ref{AI})  we set $\bo X = \bo Z$ without any loss of generality.

All  computations were performed using R 4.2.1 \citep{R} with the packages JADE \citep{MiettinenNordhausenTaskinen2017}, fICA \citep{MiettinenNordhausenTaskinen2018}, FastICA \citep{FastICA}, steadyICA \citep{steadyICA}, mvtnorm \citep{mvtnorm}, and copula \citep{copula}. The code for the competing test of \citet{MattesonTsay2017} was kindly provided by David Matteson; we used it exactly as recommended in the original paper, based on 200 permutations,   with the built-in option for FastICA, which also uses $\mathsf{tanh}$. All simulation results are based on 1,000 replications. For our tests, we used the warp-speed bootstrapping  \citep{GiacominiPolitisWhite2013}  (bootstrap sample size~1,000) and we applied the same principle also in the permutational case  ($M=1,000$ permutations)---call this {\it warp-speed permutation}. In
this warp-speed framework, the number of bootstrap samples is the number
of replications in the simulations. More discussion on the number of
bootstrap samples is in Appendix~\ref{AppB4}.

For our   tests, we considered the following test statistics, where superscripts  $\text{\rm Boot}$ and~$\text{\rm Perm}$ stand for  bootstrapped   and  permutational   critical values, respectively: 
$T_{n,G}^{\text{\rm Boot}}$ and~$T_{n,G}^{\text{\rm Perm}}$ with  Gaussian weights with $\gamma=1$; $T_{n,L}^{\text{\rm Boot}}$ and $T_{n,L}^{\text{\rm Perm}}$ with Laplace weights with~$\gamma=1$; $\tenq{T}_{n,\text{Id},G}^{\text{\rm Boot}}$ and~$\tenq{T}_{n,\text{Id},G}^{\text{\rm Perm}}$ with identity (aka Wilcoxon) scores and Gaussian weights;~$\tenq{T}_{n,\text{Id},G}^{\text{\rm Boot}}$ and~$\tenq{T}_{n,\text{Id},G}^{\text{\rm Perm}}$ with identity (aka Wilcoxon) scores and Laplace weights; $\tenq{T}_{n,\text{vdW},G}^{\text{\rm Boot}}$ and $\tenq{T}_{n,\text{vdW},G}^{\text{\rm Perm}}$ for van der Waerden scores and Gaussian weights;  $T_{n,\text{DC}}$ stands  for the test statistic of \citet{MattesonTsay2017}.

\subsection{Results for Setting 1} 

Two important preliminary questions are:  Do the tests keep the nominal $\alpha$ level? More particularly, how does the answer to this question depend on the choice of an ICA method?  Simulations under Setting 1 (which satisfies the ICM assumptions) were conducted to answer these questions;  we also included in these simulations the ``oracle tests''  based on the true values of the ${\boldsymbol Z}$'s. Note that it is well documented that ICA methods have convergence issues if sample sizes are small:  a good estimation of the latent ${\boldsymbol Z}$'s often require thousands of observations   (see \citet{MiettineNordhausenOjaTaskinen2014,MiettinenTaskinenNordhausenOja2015}).

Table~\ref{PowerH0gauss}  provides the rejection frequencies for  $T_{n,G}^{\text{\rm Boot}}$, $T_{n,G}^{\text{\rm Perm}}$,  and~$T_{n,\text{DC}}$ and the rank-based~$\tenq{T}_{n,Id,G}^{\text{\rm Boot}}$, and~$\tenq{T}_{n,Id, G}^{\text{\rm Perm}}$
 under Setting~1 for various sample sizes and various ICA estimation methods.
\begin{table}[t!]
\centering
\footnotesize
\begin{tabular*}{\textwidth}{@{\extracolsep{\fill}}*{10}{c}}
  \hline
  &  \multicolumn{4}{c}{$T_{n,G}^{\text{\rm Boot}}$} &  \multicolumn{4}{c}{$T_{n,G}^{\text{\rm Perm}}$} & $T_{n,\text{DC}}$ \\ \cline{2-5} \cline{6-9}  \cline{10-10}
$n$ & JADE & FOBI & FastICA & TRUE & JADE & FOBI & FastICA & TRUE & FastICA\\ 
  \hline
500 & 0.0700 & 0.1400 & 0.0602 & 0.0640 & 0.0590 & 0.1310 & 0.0453 & 0.0580 & 0.058\\ 
  1,000 & 0.0540 & 0.1350 & 0.0400 & 0.0640 & 0.0690 & 0.1260 & 0.0500 & 0.0470 & 0.063\\ 
  2,000 & 0.0600 & 0.1040 & 0.0590 & 0.0750 & 0.0560 & 0.1180 & 0.0510 & 0.0750 & 0.049\\ 
  4,000 & 0.0630 & 0.0910 & 0.0520 & 0.0570 & 0.0730 & 0.0840 & 0.0430 & 0.0530 & 0.053\\ 
  8,000 & 0.0700 & 0.0650 & 0.0660 & 0.0680 & 0.0700 & 0.0540 & 0.0610 & 0.0380 & 0.039\\ 
  16,000 & 0.0500 & 0.0390 & 0.0420 & 0.0550 & 0.0500 & 0.0300 & 0.0480 & 0.0670 & 0.047 \\ 
  \hline
  &  \multicolumn{4}{c}{$\tenq{T}_{n,Id,G}^{\text{\rm Boot}}$} &  \multicolumn{4}{c}{$\tenq{T}_{n,Id, G}^{\text{\rm Perm}}$}  \\  \cline{2-5}\cline{6-9}
%n & JADE & FOBI & FastICA & TRUE & Jade & Fobi & FastICA & TRUE\\ 
  
500 & 0.0930 & 0.1680 & 0.0822 & 0.0620 & 0.0910 & 0.1870 & 0.0802 & 0.0560 \\ 
  1,000 & 0.0680 & 0.1610 & 0.0390 & 0.0520 & 0.0710 & 0.1800 & 0.0580 & 0.0630 \\ 
  2,000 & 0.0810 & 0.1180 & 0.0570 & 0.0420 & 0.0750 & 0.1000 & 0.0490 & 0.0460 \\ 
  4,000 & 0.0670 & 0.1140 & 0.0460 & 0.0510 & 0.0740 & 0.1120 & 0.0550 & 0.0470 \\ 
  8,000 & 0.0530 & 0.0670 & 0.0380 & 0.0410 & 0.0710 & 0.0580 & 0.0670 & 0.0360 \\ 
  16,000 & 0.0670 & 0.0320 & 0.0500 & 0.0570 & 0.0380 & 0.0210 & 0.0590 & 0.0370 \\ 
   \hline
   \end{tabular*}
\caption{Empirical sizes (rejection frequencies over 1,000 replications) of the tests based on~$T_{n,G}^{\text{\rm Boot}}$, $T_{n,G}^{\text{\rm Perm}}$, $T_{n,\text{DC}}$, $\tenq{T}_{n,Id,G}^{\text{\rm Boot}}$, and~$\tenq{T}_{n,Id, G}^{\text{\rm Perm}}$  (nominal size $\alpha = 5\%$) under Setting 1 and 
  various ICA methods. Column TRUE corresponds to the oracle test based on   the actual  values of the $\boldsymbol Z$'s.} \vspace{-0mm}\label{PowerH0gauss}
\end{table}
The table clearly shows that the choice of an ICA method has a sizeable impact on the actual size of the tests; FastICA appears to be best from that perspective, while FOBI is worst and clearly cannot be trusted even for samples as large as  $n=4,000$. The differences between permutation and bootstrap variants are minimal and, starting from $n=2,000$ observations, the results (except for FOBI) seem satisfactory. The size of the rank-based tests is not much closer to the nominal level than that of their ``parametric'' counterparts, confirming the well-known fact that these tests (even the rank-based ones rely on an ICA step), just as ICA itself, are large-sample procedures. 
The same conclusions hold for other weight functions: see Appendix~\ref{AppB1} for the corresponding tables. 

\subsection{Results for Settings 2 and 3} 

Simulations under Settings 2 and 3  provide results on finite-sample powers (with sample sizes~$n=500, 1,000, 2,000$,  and  $4,000$). Under Setting 2, the only distribution satis\-fying the ICM assumptions is the spherical Gaussian one ($t_\infty$);  all others yield dependencies.  Also note that if the degrees of freedom are smaller than df=5,   the ICA methods considered below should not work, since they require finite 4th-order moments. 

\begin{figure}
    \centering
    \includegraphics[width=0.8\textwidth]{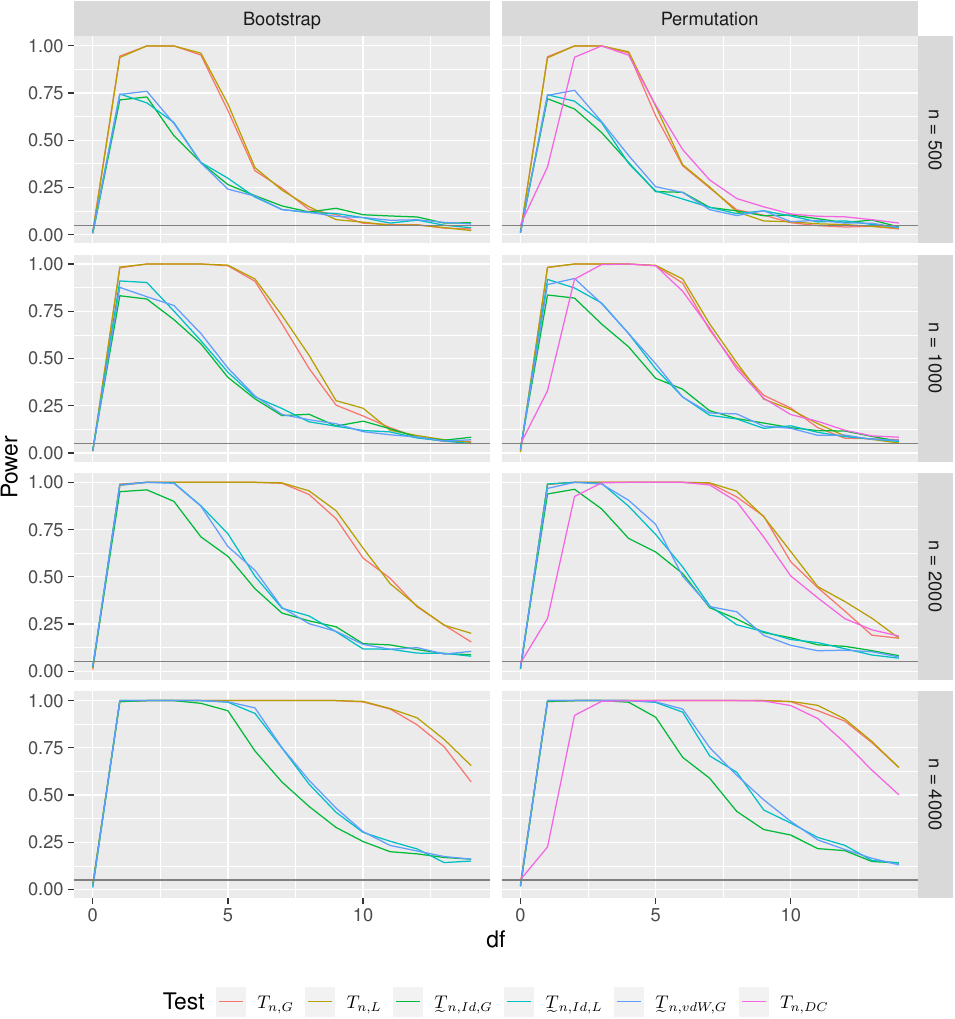}\vspace{-5mm} 
    \caption{\small Empirical powers (rejection frequencies over 1,000 replications) of the tests based on~$T_{n,G}^{\text{\rm Boot}}$,  $T_{n,L}^{\text{\rm Boot}}$,  $\tenq{T}_{n,\text{Id},G}^{\text{\rm Boot}}$,   $\tenq{T}_{n,\text{Id},L}^{\text{\rm Boot}}$, $\tenq{T}_{n,\text{vdW},G}^{\text{\rm Boot}}$, $T_{n,G}^{\text{\rm Perm}}$,  $T_{n,L}^{\text{\rm  Perm}}$,  $\tenq{T}_{n,\text{Id},G}^{\text{\rm  Perm}}$,   $\tenq{T}_{n,\text{Id},L}^{\text{\rm  Perm}}$, $\tenq{T}_{n,\text{vdW},G}^{\text{\rm Perm}}$, and~$T_{n,\text{DC}}$
       for sample sizes $n=500,\, 1,000,\, 2,000$, and $4,000$ in Setting 2 (spherical $t$), as functions of the degrees of freedom when using FastICA with bootstrap and permutational critical values, respectively. The grey horizontal line represents the nominal size $\alpha = 5\%$. Note that in the figure the Gaussian distribution ($t_\infty$) is represented under df=0.}
    \label{fig:PowerS2FastICA}
\end{figure}

Given the findings of the previous section, we only display the results based on FastICA,  postponing the FOBI and JADE   figures to  Appendix~\ref{AppB2}. In this regard, an inspection of Figure~\ref{fig:PowerS2FastICA} reveals that the Gaussian model ($t_\infty$), quite correctly, is not rejected as an~ICM. Although FastICA is not expected to work well,  the tests correctly reject the null hypothesis under non-Gaussian $\boldsymbol Z$'s for small degrees of freedom, with the rank-based procedures outperforming all other procedures. This robustness to heavy tails has been also observed by \citet{ChenBickel2005}, Section IV. 
As expected, the more degrees of freedom (hence, the closer to Gaussianity), the less powerful all tests.    
 With  Gaussian or Laplace weights, in this setting, our tests are doing equally well as~$T_{n,\text{DC}}$ but the power of the rank-based tests decreases more sharply. Notice that Figure~\ref{fig:PowerS2FastICA} 
  omits results for larger degrees of freedom as all powers then tend to the nominal level $\alpha = 5\%$. Similar figures, with similar conclusions, 
  are obtained for JADE  methods, while FOBI is not working at all: see  Appendix~\ref{AppB2}. 
  
\begin{figure}
    \centering
    \includegraphics[width=0.8\textwidth]{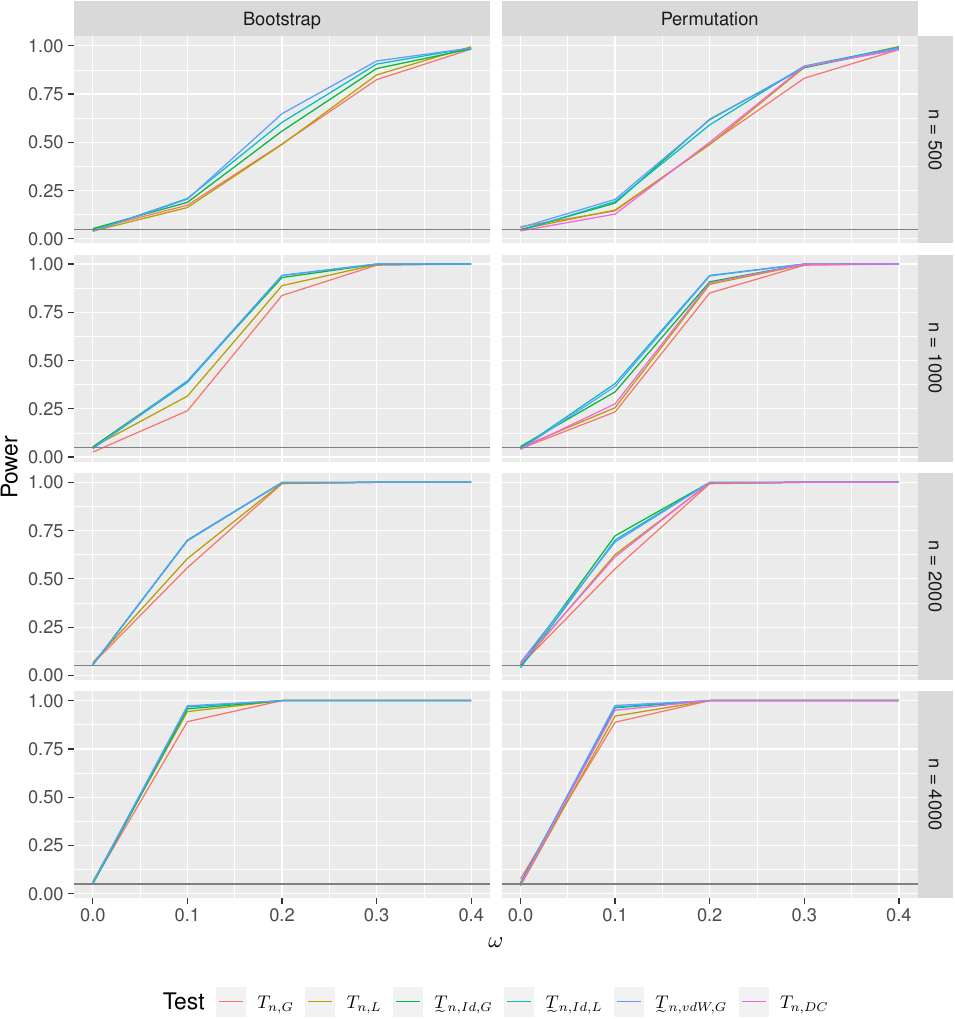}\vspace{-5mm} 
   \caption{\small Empirical powers (rejection frequencies over 1,000 replications) of the tests based on $T_{n,G}^{\text{\rm Boot}}$,  $T_{n,L}^{\text{\rm Boot}}$,  $\tenq{T}_{n,\text{Id},G}^{\text{\rm Boot}}$,   $\tenq{T}_{n,\text{Id},L}^{\text{\rm Boot}}$, $\tenq{T}_{n,\text{vdW},G}^{\text{\rm Boot}}$, $T_{n,G}^{\text{\rm Perm}}$,  $T_{n,L}^{\text{\rm  Perm}}$,  $\tenq{T}_{n,\text{Id},G}^{\text{\rm  Perm}}$,   $\tenq{T}_{n,\text{Id},L}^{\text{\rm  Perm}}$, $\tenq{T}_{n,\text{vdW},G}^{\text{\rm Perm}}$, and~$T_{n,\text{DC}}$
       for sample sizes $n=500,\, 1,000,\, 2,000$, and $4,000$ in Setting 3 (Clayton copula), as  functions of the copula  parameter $\omega$  when using FastICA and bootstrap or permutational critical values, respectively. The grey horizontal line represents the nominal size $\alpha = 5\%$.}
 \label{fig:PowerS3FastICA}
\end{figure}

Similarly, under Setting~3, the only independent component model is obtained for~$\omega=0$ and the dependence between the components increases with $\omega$. Figure~\ref{fig:PowerS3FastICA}  clearly shows, for the FastICA method, that the rejection frequency matches the nominal $\alpha = 5\%$ level at the ICM (for $\omega =~\!0$). Then, with increasing~$\omega$, the power increases, and, with larger sample sizes, even weaker dependencies are well detected. The rank-based tests are outperforming all other ones, including the competing test based on~$T_{n,\text{DC}}$. Again, larger values of $\omega$ are omitted in the figure as the power there is trivially equal to one. The corresponding figures for the FOBI- and  JADE-based methods are provided in Appendix~\ref{AppB3},  indicating, as before, that FOBI should not be used while JADE seems as good as FastICA.

In all three settings,  the bootstrap and permutation versions yield essentially identical performances. 
  Which one is to be 
 adopted, therefore, does not matter much (for a discussion of the number of bootstrap samples/permutations, see Appendix~\ref{AppB4}).  
 
 More importantly, the ICA method involved needs to work reasonably well in the estimation of the mixing matrix $\boldsymbol\Omega$, hence the latent~$\boldsymbol Z$'s. It is a well-known fact, in the ICA literature, that this requires quite large samples, and as a result, so do our tests. This is hardly an issue since  ICA is never considered in the analysis of small or moderate samples. This also might be the main reason why FOBI, which is particularly greedy in that respect,  is working poorly here. We thus recommend JADE or FastICA. 
% had here often convergence issues. 
Note that  FastICA was also used by \citet{MattesonTsay2017}  in the implementation of the competing test based on~$T_{n,\text{DC}}$. It might be worthwhile to consider also less classical  ICA methods such as, for example, the R-estimator of  \citet{HallinMehta2015}. Such methods, however, usually come with a higher computational burden, which makes them impractical in the bootstrap or sampling framework used here. Different score functions $\bf J$ also are more suitable at different distributions;  based on our results, it seems that a Gaussian weight function (using either the original observations or their ranks) is a good choice.

\section{A real-data application}\label{sec_8}

ICA is as a widely employed tool in biomedical signal processing. In this context, we will examine an electrocardiogram (ECG) dataset featuring eight sensors positioned on a pregnant woman's skin (for comprehensive details and data availability, see  
\citet{deLathauwerdeMoorVandewalle2000} or
 \citet{MiettinenNordhausenTaskinen2017}). ECG records the skin's electrical potential resulting from muscle activities, with the primary objective of measuring the fetal heartbeat. However, the ECG recording also captures the mother's heartbeat, as well as some other muscle activities collectively referred to as {\it artifacts}. ICA frequently finds application as a tool for artifact removal \citep{HeCliffordTarassenko:2006}, with artifacts being treated as independent disturbances.

\begin{figure}[b!]
    \centering
    \includegraphics[width=0.8\textwidth]{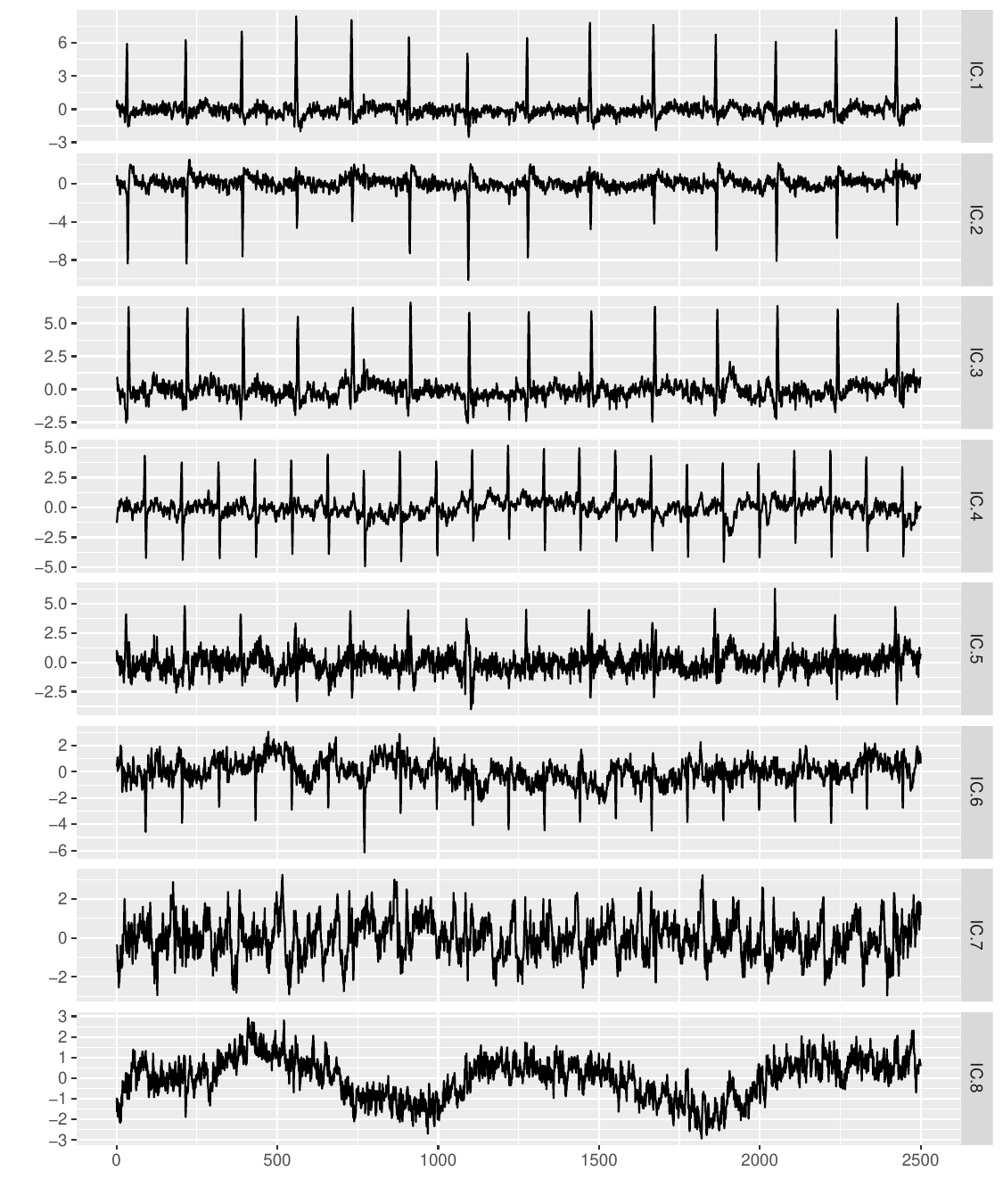}\vspace{-5mm} 
    \caption{Independent components based on JADE for the ECG measurements of the pregnant women dataset.}
    \label{fig:ECGjade}
\end{figure}

The original dataset is presented in Figure~\ref{fig:ECGdata} (Appendix~\ref{AppC}). Figure~\ref{fig:ECGjade} below displays the estimated independent components based on JADE. 
The fourth independent component, according to experts,  corresponds to the fetal heartbeat, while the last two %components
 likely are artifacts. Other components seem to be associated with the mother's heartbeat. To facilitate our tests, we initially removed second-order serial dependencies from the estimated components. We accomplished this by fitting an autoregressive (AR) model to each component, with the order selection performed via   Akaike's criterion (AIC). The resulting eight residual series are depicted in  Appendix~\ref{AppC},   Figure~\ref{fig:ECGres}.

Subsequently, we employed 500 bootstrap samples to subject the resulting vectors of residuals to hypothesis testing using the test statistics $T_{n,G}^{\text{\rm Boot}}$ and $\tenq{T}_{n, \text{Id}, G}^{\text{\rm Boot}}$ to assess if they satisfy the  ICM assumptions. In both cases, we obtained a  0.002  $p$--value,  
which leads to a rejection of the null hypothesis (a result that might be due to the fact  that the mother's heartbeat is present in several components).  The ICA analysis of this dataset, thus, should be taken with great care. 

We then restricted the analysis to the subvector identified by the experts as a vector of artifacts. Performing the same tests as above yields  $p$--values  0.992 and 0.936, respectively,
in agreement with the fact that the artifacts originate from independent sources.  

\section{Conclusions}\label{sec_9} 

 Affine--invariant and globally consistent CF--based tests for the validity of the ICM are considered. Due to the complexity of the large-sample distribution of the test statistics, the tests are applied in their permutation or bootstrap versions using either the original observations or the corresponding ranks, and are implemented with a variety of estimation methods of the mixing matrix, and with suitable weight functions that lead to convenient closed-form expressions. Based on numerical results for the suggested tests we wish to register the following warning messages against casually applying the ICM: (1) large sample sizes are necessary, even in the thousands, for proper estimation and model fitting; (2) in most typical ICA applications, {\it{deserialization}} procedures for the estimated independent components are required to remove time series--type dependencies. 
With regards to the Monte Carlo behavior of the tests we point out the following:  (1) the type of estimation does affect the empirical size of the test with FastICA being more reliable, followed by JADE; (2) rejection under alternatives reaches reasonable levels and is sample-size increasing, in line with consistency, with original and rank-based procedures being overall comparable, whether permutation- or bootstrap-based implementations are considered;  (3) the new tests either seem to attain comparable powers or outperform their distance covariance--based counterpart, depending on the type of deviations from the null hypothesis.

\bibliographystyle{abbrvnat}
\bibliography{Refs}

\begin{thebibliography}{68}
\providecommand{\natexlab}[1]{#1}
\providecommand{\url}[1]{\texttt{#1}}
\expandafter\ifx\csname urlstyle\endcsname\relax
  \providecommand{\doi}[1]{doi: #1}\else
  \providecommand{\doi}{doi: \begingroup \urlstyle{rm}\Url}\fi

\bibitem[Blum et~al.(1961)Blum, Kiefer, and
  Rosenblatt]{BlumKieferRosenblatt1961}
J.~R. Blum, J.~Kiefer, and M.~Rosenblatt.
\newblock {Distribution-free tests of independence based on the sample
  distribution function}.
\newblock \emph{Annals of Mathematical Statistics}, 32:\penalty0 485--498,
  1961.

\bibitem[Bogachev(2018)]{Bogachev2018}
V.~I. Bogachev.
\newblock \emph{Weak Convergence of Measures}.
\newblock American Mathematical Society, Providence, 2018.

\bibitem[Bosq(2000)]{Bosq2000}
D.~Bosq.
\newblock \emph{Linear Processes in Function Spaces: Theory and Applications}.
\newblock Springer, New York, 2000.

\bibitem[Cardoso(1989)]{Cardoso1989}
J.-F. Cardoso.
\newblock Source separation using higher order moments.
\newblock In \emph{Proceedings of the IEEE International Conference on
  Acoustics, Speech and Signal Processing}, pages 2109--2112. IEEE, 1989.

\bibitem[Cardoso and Souloumiac(1993)]{CardosoSouloumiac1993}
J.-F. Cardoso and A.~Souloumiac.
\newblock Blind beamforming for non-{G}aussian signals.
\newblock In \emph{{IEE} Proceedings {F}-Radar and Signal Processing}, volume
  140, pages 362--370, 1993.

\bibitem[Chakraborty and Zhang(2019)]{ChakrabortyZhang2019}
S.~Chakraborty and X.~Zhang.
\newblock Distance metrics for measuring joint dependence with application to
  causal inference.
\newblock \emph{Journal of the American Statistical Association}, 114:\penalty0
  1638--1650, 2019.

\bibitem[Chen and Bickel(2005)]{ChenBickel2005}
A.~Chen and P.~Bickel.
\newblock Consistent independent component analysis and prewhitening.
\newblock \emph{IEEE Transactions on Signal Processing}, 53:\penalty0
  3625--3632, 2005.

\bibitem[Chen et~al.(2019)Chen, Meintanis, and Zhu]{ChenMeintanisZhu2019}
F.~Chen, S.~G. Meintanis, and L.~Zhu.
\newblock On some characterizations and multidimensional criteria for testing
  homogeneity, symmetry and independence.
\newblock \emph{Journal of Multivariate Analysis}, 173:\penalty0 125--144,
  2019.

\bibitem[Chernozhukov et~al.(2017)Chernozhukov, Galichon, Hallin, and
  Henry]{ChernozhukovGalichonHallinHenry2017}
V.~Chernozhukov, A.~Galichon, M.~Hallin, and M.~Henry.
\newblock {Monge–Kantorovich depth, quantiles, ranks and signs}.
\newblock \emph{Annals of Statistics}, 45:\penalty0 223--256, 2017.

\bibitem[Comon and Jutten(2010)]{ComonJutten2010}
P.~Comon and C.~Jutten.
\newblock \emph{Handbook of Blind Source Separation: {I}ndependent Component
  Analysis and Applications}.
\newblock Academic Press, Oxford, 2010.

\bibitem[Cs{\" {o}}rg{\H o}(1981)]{Csorgo1981}
S.~Cs{\" {o}}rg{\H o}.
\newblock Multivariate empirical characteristic functions.
\newblock \emph{Zeitschrift für Wahrscheinlichkeitstheorie und Verwandte
  Gebiete}, 55:\penalty0 203--229, 1981.

\bibitem[Cs{\" o}rg{\H o}(1985)]{Csorgo1985}
S.~Cs{\" o}rg{\H o}.
\newblock Testing for independence by the empirical characteristic function.
\newblock \emph{Journal of Multivariate Analysis}, 16:\penalty0 290--299, 1985.

\bibitem[Cs{\" o}rg{\H o} and Hall(1982)]{CsorgoHall1982}
S.~Cs{\" o}rg{\H o} and P.~Hall.
\newblock Estimable versions of {G}riffiths' measure of association.
\newblock \emph{Australian Journal of Statistics}, 24:\penalty0 296--308, 1982.

\bibitem[Cuppens(1975)]{Cuppens1975}
R.~Cuppens.
\newblock \emph{Decomposition of Multivariate Probabilities}.
\newblock Academic Press, New York, 1975.

\bibitem[de~Lathauwer et~al.(2000)de~Lathauwer, de~Moor, and
  Vandewalle]{deLathauwerdeMoorVandewalle2000}
L.~de~Lathauwer, B.~de~Moor, and J.~Vandewalle.
\newblock Fetal electrocardiogram extraction by blind source subspace
  separation.
\newblock \emph{IEEE Transactions on Biomedical Engineering}, 47:\penalty0
  567--572, 2000.

\bibitem[Deheuvels(1981)]{Deheuvels1981}
P.~Deheuvels.
\newblock An asymptotic decomposition for multivariate distribution-free tests
  of independence.
\newblock \emph{Journal of Multivariate Analysis}, 11:\penalty0 102--113, 1981.

\bibitem[Edelmann et~al.(2019)Edelmann, Fokianos, and Pitsillou]{FP19}
D.~Edelmann, K.~Fokianos, and M.~Pitsillou.
\newblock An updated literature review of distance correlation and its
  applications to time series.
\newblock \emph{International Statistical Review}, 87:\penalty0 237--262, 2019.

\bibitem[Eriksson and Koivunen(2003)]{ErikssonKoivunen2003}
J.~Eriksson and V.~Koivunen.
\newblock Characteristic-function-based independent component analysis.
\newblock \emph{Signal Processing}, 83:\penalty0 2195--2208, 2003.

\bibitem[Fan et~al.(2017)Fan, {Lafaye de Micheaux}, Penev, and
  Salopek]{FanDeMicheauxPenevSalopek2017}
Y.~Fan, P.~{Lafaye de Micheaux}, S.~Penev, and D.~Salopek.
\newblock Multivariate nonparametric test of independence.
\newblock \emph{Journal of Multivariate Analysis}, 153:\penalty0 189--210,
  2017.

\bibitem[Feuerverger(1993)]{Feuerverger1993}
A.~Feuerverger.
\newblock A consistent test for bivariate dependence.
\newblock \emph{International Statistical Review}, 61:\penalty0 419--433, 1993.

\bibitem[Garcia-Ferrer et~al.(2012)Garcia-Ferrer, Gonzalez-Prieto, and
  Pena]{GarciaFerrerGonzalezPrietoPena2012}
A.~Garcia-Ferrer, E.~Gonzalez-Prieto, and D.~Pena.
\newblock A conditionally heteroskedastic independent factor model with an
  application to financial stock returns.
\newblock \emph{International Journal of Forecasting}, 28:\penalty0 70--93,
  2012.

\bibitem[Genest et~al.(2019)Genest, Ne\v{s}lehov\'a, R\'emillard, and
  Murphy]{GenestNeslehovaRemillardMurphy2019}
C.~Genest, J.~G. Ne\v{s}lehov\'a, B.~R\'emillard, and O.~A. Murphy.
\newblock Testing for independence in arbitrary distributions.
\newblock \emph{Biometrika}, 106:\penalty0 47--68, 2019.

\bibitem[Genz et~al.(2021)Genz, Bretz, Miwa, Mi, Leisch, Scheipl, and
  Hothorn]{mvtnorm}
A.~Genz, F.~Bretz, T.~Miwa, X.~Mi, F.~Leisch, F.~Scheipl, and T.~Hothorn.
\newblock \emph{{mvtnorm}: Multivariate Normal and t Distributions}, 2021.
\newblock URL \url{https://CRAN.R-project.org/package=mvtnorm}.
\newblock R package version 1.1-3.

\bibitem[Ghosal and Sen(2022)]{GhosalSen2022}
P.~Ghosal and B.~Sen.
\newblock {Multivariate ranks and quantiles using optimal transport:
  Consistency, rates and nonparametric testing}.
\newblock \emph{Annals of Statistics}, 50:\penalty0 1012--1037, 2022.

\bibitem[Giacomini et~al.(2013)Giacomini, Politis, and
  White]{GiacominiPolitisWhite2013}
R.~Giacomini, D.~Politis, and H.~White.
\newblock {A warp-speed method for conducting Monte Carlo experiments involving
  bootstrap estimators}.
\newblock \emph{Econometric Theory}, 29:\penalty0 567--589, 2013.

\bibitem[Gourieroux et~al.(2017)Gourieroux, Monfort, and
  Renne]{GourierouxMonfortRenne2017}
C.~Gourieroux, A.~Monfort, and J.-P. Renne.
\newblock Statistical inference for independent component analysis: Application
  to structural {VAR} models.
\newblock \emph{Journal of Econometrics}, 196:\penalty0 111--126, 2017.

\bibitem[Hai(2020)]{Hai2020}
T.~Hai.
\newblock Estimation of volatility causality in structural autoregressions with
  heteroskedasticity using independent component analysis.
\newblock \emph{Statistical Papers}, 61:\penalty0 1--16, 2020.

\bibitem[H{\'a}jek and {\v{S}}id{\'a}k(1967)]{hajekSidak1967}
J.~H{\'a}jek and Z.~{\v{S}}id{\'a}k.
\newblock \emph{Theory of Rank Tests}.
\newblock Academic Press, New York, 1967.

\bibitem[Hallin and Mehta(2015)]{HallinMehta2015}
M.~Hallin and C.~Mehta.
\newblock R-estimation for asymmetric independent component analysis.
\newblock \emph{Journal of the American Statistical Association}, 110:\penalty0
  218--232, 2015.

\bibitem[Hallin et~al.(2021)Hallin, del Barrio, Cuesta-Albertos, and
  Matr{\'a}n]{HallinBarrioCuestaAlbertosMatran2021}
M.~Hallin, E.~del Barrio, J.~Cuesta-Albertos, and C.~Matr{\'a}n.
\newblock {Distribution and quantile functions, ranks and signs in dimension d:
  A measure transportation approach}.
\newblock \emph{Annals of Statistics}, 49:\penalty0 1139--1165, 2021.

\bibitem[He et~al.(2006)He, Clifford, and
  Tarassenko]{HeCliffordTarassenko:2006}
T.~He, G.~Clifford, and L.~Tarassenko.
\newblock Application of independent component analysis in removing artefacts
  from the electrocardiogram.
\newblock \emph{Neural Computing \& Applications}, 15:\penalty0 105--116, 2006.

\bibitem[Henze et~al.(2014)Henze, Hl\'avka, and
  Meintanis]{HenzeHlavkaMeintanis2014}
N.~Henze, Z.~Hl\'avka, and S.~Meintanis.
\newblock Testing for spherical symmetry via the empirical characteristic
  function.
\newblock \emph{Statistics}, 48:\penalty0 1282--1296, 2014.

\bibitem[Herwartz and Maxand(2020)]{HerwartzMaxand2020}
H.~Herwartz and S.~Maxand.
\newblock Nonparametric tests for independence: a review and comparative
  simulation study with an application to malnutrition in {India}.
\newblock \emph{Statistical Papers}, 61:\penalty0 2175--2201, 2020.

\bibitem[Hofert et~al.(2023)Hofert, Kojadinovic, Maechler, and Yan]{copula}
M.~Hofert, I.~Kojadinovic, M.~Maechler, and J.~Yan.
\newblock \emph{copula: Multivariate Dependence with Copulas}, 2023.
\newblock URL \url{https://CRAN.R-project.org/package=copula}.

\bibitem[Hyvarinen(1999)]{Hyvarinen1999}
A.~Hyvarinen.
\newblock Fast and robust fixed-point algorithms for independent component
  analysis.
\newblock \emph{IEEE Transactions on Neural Networks}, 10:\penalty0 626--634,
  1999.

\bibitem[Hyvarinen and Oja(1997)]{HyvarinenOja1997}
A.~Hyvarinen and E.~Oja.
\newblock A fast fixed-point algorithm for independent component analysis.
\newblock \emph{Neural Computation}, 9:\penalty0 1483--1492, 1997.

\bibitem[Ilmonen and Paindaveine(2011)]{IlmonenPaindaveine2011}
P.~Ilmonen and D.~Paindaveine.
\newblock {Semiparametrically efficient inference based on signed ranks in
  symmetric independent component models}.
\newblock \emph{Annals of Statistics}, 39:\penalty0 2448--2476, 2011.

\bibitem[Kankainen(1995)]{Kankainen1995}
A.~Kankainen.
\newblock \emph{Consistent testing of total independence based on the empirical
  characteristic function}.
\newblock PhD thesis, University of Jyv\"askyl\"a, 1995.

\bibitem[Kankainen and Ushakov(1998)]{KankainenUshakov1998}
A.~Kankainen and N.~Ushakov.
\newblock A consistent modification of a test for independence based on the
  empirical characteristic function.
\newblock \emph{Journal of Mathematical Sciences}, 89:\penalty0 1486--1494,
  1998.

\bibitem[Kozubowski et~al.(2013)Kozubowski, Podg\'orski, and
  Rychlik]{KozubowskiPodgorskiRychlik2013}
T.~J. Kozubowski, K.~Podg\'orski, and I.~Rychlik.
\newblock Multivariate generalized {L}aplace distribution and related random
  fields.
\newblock \emph{Journal of Multivariate Analysis}, 113:\penalty0 59--72, 2013.

\bibitem[Lukacs(1970)]{Lukacs1970}
E.~Lukacs.
\newblock \emph{Characteristic Functions}.
\newblock Hafner Publishing Company, London, 1970.

\bibitem[Marchini et~al.(2021)Marchini, Heaton, and Ripley]{FastICA}
J.~L. Marchini, C.~Heaton, and B.~D. Ripley.
\newblock \emph{fastICA:\! FastICA~Algo\-rithms to Perform ICA and Projection
  Pursuit}, 2021.
\newblock URL \url{https://CRAN.R-project.org/package=fastICA}.

\bibitem[Matteson and Tsay(2017)]{MattesonTsay2017}
D.~S. Matteson and R.~S. Tsay.
\newblock Independent component analysis via distance covariance.
\newblock \emph{Journal of the American Statistical Association}, 112:\penalty0
  623--637, 2017.

\bibitem[Meintanis and Iliopoulos(2008)]{MeintanisIliopoulos2008}
S.~G. Meintanis and G.~Iliopoulos.
\newblock Fourier methods for testing multivariate independence.
\newblock \emph{Computational Statistics \& Data Analysis}, 52:\penalty0
  1884--1895, 2008.

\bibitem[Miettinen et~al.(2014)Miettinen, Nordhausen, Oja, and
  Taskinen]{MiettineNordhausenOjaTaskinen2014}
J.~Miettinen, K.~Nordhausen, H.~Oja, and S.~Taskinen.
\newblock Deflation-based {FastICA} with adaptive choices of nonlinearities.
\newblock \emph{IEEE Transactions on Signal Processing}, 62:\penalty0
  5716--5724, 2014.

\bibitem[Miettinen et~al.(2015)Miettinen, Taskinen, Nordhausen, and
  Oja]{MiettinenTaskinenNordhausenOja2015}
J.~Miettinen, S.~Taskinen, K.~Nordhausen, and H.~Oja.
\newblock Fourth moments and independent component analysis.
\newblock \emph{Statistical Science}, 30:\penalty0 372--390, 2015.

\bibitem[Miettinen et~al.(2017{\natexlab{a}})Miettinen, Nordhausen, Oja,
  Taskinen, and Virta]{MiettinenNordhausenOjaTaskinenVirta2017}
J.~Miettinen, K.~Nordhausen, H.~Oja, S.~Taskinen, and J.~Virta.
\newblock The squared symmetric {FastICA} estimator.
\newblock \emph{Signal Processing}, 131:\penalty0 402--411, 2017{\natexlab{a}}.

\bibitem[Miettinen et~al.(2017{\natexlab{b}})Miettinen, Nordhausen, and
  Taskinen]{MiettinenNordhausenTaskinen2017}
J.~Miettinen, K.~Nordhausen, and S.~Taskinen.
\newblock Blind source separation based on joint diagonalization in \textsf{R}:
  The packages \texttt{JADE} and \texttt{BSSasymp}.
\newblock \emph{Journal of Statistical Software}, 76:\penalty0 1--31,
  2017{\natexlab{b}}.

\bibitem[Miettinen et~al.(2018)Miettinen, Nordhausen, and
  Taskinen]{MiettinenNordhausenTaskinen2018}
J.~Miettinen, K.~Nordhausen, and S.~Taskinen.
\newblock {fICA}: {FastICA} algorithms and their improved variants.
\newblock \emph{The R Journal}, 10:\penalty0 148--158, 2018.

\bibitem[Miettinen et~al.(2020)Miettinen, Matilainen, Nordhausen, and
  Taskinen]{MiettinenMatilainenNordhausenTaskinen2020}
J.~Miettinen, M.~Matilainen, K.~Nordhausen, and S.~Taskinen.
\newblock Extracting conditionally heteroskedastic components using independent
  component analysis.
\newblock \emph{Journal of Time Series Analysis}, 41:\penalty0 293--311, 2020.

\bibitem[Nolan(2013)]{Nolan2013}
J.~Nolan.
\newblock Multivariate elliptically contoured stable distributions: theory and
  estimation.
\newblock \emph{Computational Statistics}, 28:\penalty0 2067--2089, 2013.

\bibitem[Nordhausen and Oja(2018)]{NordhausenOja2018}
K.~Nordhausen and H.~Oja.
\newblock Independent component analysis: A statistical perspective.
\newblock \emph{WIREs: Computational Statistics}, 10:\penalty0 e1440, 2018.

\bibitem[Oja et~al.(2016)Oja, Paindaveine, and
  Taskinen]{OjaPaindaveineTaskinen2016}
H.~Oja, D.~Paindaveine, and S.~Taskinen.
\newblock {Affine-invariant rank tests for multivariate independence in
  independent component models}.
\newblock \emph{Electronic Journal of Statistics}, 10:\penalty0 2372--2419,
  2016.

\bibitem[Pfister et~al.(2017)Pfister, B\"uhlmann, Sch\"olkopf, and
  Peters]{PfisterBuhlmannScholkopfPeters2017}
N.~Pfister, P.~B\"uhlmann, B.~Sch\"olkopf, and J.~Peters.
\newblock Kernel-based tests for joint independence.
\newblock \emph{Journal of the Royal Statistical Society Series B: Statistical
  Methodology}, 80:\penalty0 5--31, 2017.

\bibitem[{R Core Team}(2023)]{R}
{R Core Team}.
\newblock \emph{R: A Language and Environment for Statistical Computing}.
\newblock R Foundation for Statistical Computing, Vienna, Austria, 2023.
\newblock URL \url{https://www.R-project.org/}.

\bibitem[Risk et~al.(2015)Risk, James, and Matteson]{steadyICA}
B.~B. Risk, N.~A. James, and D.~S. Matteson.
\newblock \emph{steadyICA: ICA and Tests of Independence via Multivariate
  Distance Covariance}, 2015.
\newblock URL \url{https://CRAN.R-project.org/package=steadyICA}.
\newblock R package version 1.0.

\bibitem[Rossberg(1995)]{Rossberg1995}
H.~Rossberg.
\newblock Positive definite probability densities and probability
  distributions.
\newblock \emph{Journal of Mathematical Sciences}, 76:\penalty0 2181--2197,
  1995.

\bibitem[Roy et~al.(2020)Roy, Sarkar, Ghosh, and
  Goswami]{RoySarkarGhoshGoswami2020}
A.~Roy, S.~Sarkar, A.~K. Ghosh, and A.~Goswami.
\newblock On some consistent tests of mutual independence among several random
  vectors of arbitrary dimensions.
\newblock \emph{Statistics and Computing}, 30:\penalty0 1707--1723, 2020.

\bibitem[Sejdinovic et~al.(2013)Sejdinovic, Sriperumbudur, Gretton, and
  Fukumizu]{Sejd13}
D.~Sejdinovic, B.~Sriperumbudur, A.~Gretton, and K.~Fukumizu.
\newblock Equivalence of distance-based and {RKHS}--based statistics in
  hypothesis testing.
\newblock \emph{Annals of Statistics}, 41:\penalty0 2263--2291, 2013.

\bibitem[Shi et~al.(2022{\natexlab{a}})Shi, Drton, and Han]{ShiDrtonHan2022}
H.~Shi, M.~Drton, and F.~Han.
\newblock Distribution-free consistent independence tests via center-outward
  ranks and signs.
\newblock \emph{Journal of the American Statistical Association}, 117:\penalty0
  395--410, 2022{\natexlab{a}}.

\bibitem[Shi et~al.(2022{\natexlab{b}})Shi, Hallin, Drton, and
  Han]{ShiHallinDrtonHan2022}
H.~Shi, M.~Hallin, M.~Drton, and F.~Han.
\newblock {On universally consistent and fully distribution-free rank tests of
  vector independence}.
\newblock \emph{Annals of Statistics}, 50:\penalty0 1933--1959,
  2022{\natexlab{b}}.

\bibitem[Shi et~al.(2023)Shi, Drton, Hallin, and Han]{ShiHallinDrtonHan2023}
H.~Shi, M.~Drton, M.~Hallin, and F.~Han.
\newblock Distribution-free tests of multivariate independence based on
  center-outward quadrant, {S}pearman, and {K}endall statistics.
\newblock \emph{Bernoulli, {\rm to appear}}, 2023.

\bibitem[Sz{\'e}kely and Rizzo(2023)]{SR23}
G.~J. Sz{\'e}kely and M.~L. Rizzo.
\newblock \emph{The Energy of Data and Distance Correlation}.
\newblock CRC Press, Florida, USA, 2023.

\bibitem[Sz{\'e}kely et~al.(2007)Sz{\'e}kely, Rizzo, and
  Bakirov]{SzekelyRizzoBakirov2007}
G.~J. Sz{\'e}kely, M.~L. Rizzo, and N.~K. Bakirov.
\newblock {Measuring and testing dependence by correlation of distances}.
\newblock \emph{Annals of Statistics}, 35:\penalty0 2769--2794, 2007.

\bibitem[Theis et~al.(2011)Theis, Kawanabe, and
  Muller]{TheisKawanabeMuller2011}
F.~J. Theis, M.~Kawanabe, and K.-R. Muller.
\newblock Uniqueness of non-{G}aussianity-based dimension reduction.
\newblock \emph{IEEE Transactions on Signal Processing}, 59:\penalty0
  4478--4482, 2011.

\bibitem[Virta and Nordhausen(2017)]{VirtaNordhausen2017}
J.~Virta and K.~Nordhausen.
\newblock On the optimal non-linearities for {G}aussian mixtures in
  {F}ast{ICA}.
\newblock In P.~Tichavsk{\'y}, M.~Babaie-Zadeh, O.~J. Michel, and
  N.~Thirion-Moreau, editors, \emph{Latent Variable Analysis and Signal
  Separation}, pages 427--437. Springer, 2017.

\bibitem[Wei(2014)]{Wei2014}
T.~Wei.
\newblock On the spurious solutions of the {F}ast{ICA} algorithm.
\newblock In \emph{2014 IEEE Workshop on Statistical Signal Processing (SSP)},
  pages 161--164, 2014.

\bibitem[Wilks(1935)]{Wilks1935}
S.~S. Wilks.
\newblock On the independence of $k$ sets of normally distributed statistical
  variables.
\newblock \emph{Econometrica}, 3:\penalty0 309--326, 1935.

\end{thebibliography}

\newpage

\noindent \Large{\bf Appendix}\bigskip

\appendix
\renewcommand\thefigure{A.\arabic{figure}}    
\setcounter{figure}{0}  

% Change table labelling
\renewcommand\thetable{A.\arabic{table}}    
\setcounter{table}{0} 

\spacingset{1.9} % DON'T change the spacing!

\section{On the asymptotic behavior of the empirical independence process}  \label{AppA}
%\subsection{Asymptotic behavior under the null hypothesis}
%In this section, we discuss the asymptotic distribution of $T_{n,W}$ and $S_n$ under the null hypothesis ${\cal{H}}_0$. To this end, 
Letting   $\delta^{(n)}(\boldsymbol t)\coloneqq \sqrt{n}D_n(\boldsymbol t)\coloneqq \delta^{(n)}_{\Re}(\boldsymbol t)+\ii \: \delta^{(n)}_{\Im}(\boldsymbol t)$, with $D_n(\cdot)$  defined in (\ref{D}),  call  {\it empirical independence process} the collection  $\{\delta^{(n)}(\boldsymbol t)\vert \boldsymbol t\in{\mathbb R}^p\}$. In this appendix, we study the behavior as $n\to\infty$ of~$\{\delta^n(\boldsymbol t)\vert \boldsymbol t\}$ under the null hypothesis. 
In doing so we illustrate the fact that, since    $T_{n,W}$ is a  functional of the empirical independent process, the respective limit null distribution, as we shall see, depends on the unspecified distributions of the independent components of the vector $\boldsymbol Z=(Z_1,...,Z_p)^\top$, and the estimation method employed for the mixing matrix $\boldsymbol \Omega$. 
Due to this dependence on unspecified nuisances, this asymptotic distribution is hardly tractable and thus inappropriate for computing critical values. This is why we privilege bootstrapped or permutational critical values or distribution-free rank-based test statistics in the practical implementation of our tests. 

Let $\phi_n(\boldsymbol t)\coloneqq n^{-1} \sum_{j=1}^n {\rm{e}}^{\imath \boldsymbol t^\top {\bf {X}}_j}$  denote the empirical CF computed from  the $n$  i.i.d.\ copies~$({\bf {X}}_j, \ j=1,...,n)$ of the arbitrary random vector $\bf X$, and write $F$ and  $\phi$ for its distribution and CFs,  respectively. It is well known, see \citet{Csorgo1985}, that, under a rather mild tail condition, the empirical
CF process $\Xi_n(\boldsymbol t):=\left\{\sqrt{n}(\phi_n(\boldsymbol t)-\phi(\boldsymbol t))\vert\, \boldsymbol t\right\}$ converges weakly  to a limit Gaussian process $\{\Xi(\boldsymbol t) \vert\,    {\boldsymbol t}\}$, over compact subsets $C$ of $\mathbb R^p$. We will also use the fact that   $\sup_{\boldsymbol t \in C}|\phi_n(\boldsymbol t)-\phi(\boldsymbol t)|\to 0$, almost surely.

Now consider the empirical independence process. Recall from \eqref{indep} the notation~$\varphi$ and~$\varphi_\ell$, respectively,   for the characteristic functions of $\boldsymbol Z$ and~$Z_\ell$,  $\ell=1,...,p$,   and note that, under the null  hypothesis ${\cal{H}}_0$   in \eqref{null}, we have 
\begin{eqnarray}   \nonumber    \delta^{(n)}(\boldsymbol t)&=&\sqrt{n} (\varphi^{(n)}\left(\boldsymbol t)-\varphi(\boldsymbol t)\right)-\sqrt{n} \left(\prod_{\ell=1}^p
\varphi^{(n)}_{\ell}(t_\ell)-\prod_{\ell=1}^p
\varphi_{\ell}(t_\ell)\right) \\ \label{A.1}
&=&
\sqrt{n} (\varphi^{(n)}\left(\boldsymbol t)-\varphi(\boldsymbol t)\right) \\ \nonumber &&-
\sum_{\ell=1}^p \sqrt{n} (\varphi^{(n)}_{\ell}(t_\ell)-\varphi_{\ell}(t_\ell))\prod_{m<\ell} 
\varphi_{m}(t_m)\prod_{m>\ell} 
\varphi^{(n)}_{m}(t_m), \end{eqnarray}
where 
the last equality follows from the algebraic identity $$\prod_{\ell=1}^p x_\ell -\prod_{\ell=1}^p y_\ell=\sum_{\ell=1}^p (x_\ell-y_\ell) \prod_{m=1}^{\ell-1} y_m   \prod_{m=\ell+1}^{p} x_m.$$ 
Recall also from \eqref{joint}--\eqref{marg}  the notation $\varphi^{(n)}$ and $\varphi^{(n)}_{\ell}$ for the joint and marginal empirical CF of $\widehat {\bo Z}$ and $\widehat Z_{\ell}$, respectively, and write  
\begin{align*}
\varphi^{(n)}({\boldsymbol{t}})&\coloneqq \varphi^{(n)}_{\Re}(\boldsymbol t)+\ii\, \varphi^{(n)}_{\Im}(\boldsymbol t), 
&\varphi({\boldsymbol{t}})&\coloneqq \varphi_\Re (\boldsymbol t)+\ii\,  \varphi_\Im (\boldsymbol t),\\  
 \varphi_{\ell}(t)&\coloneqq \varphi_{\ell\Re}(t)+\ii\, \varphi_{\ell\Im}(t),\quad\quad\quad 
\text{and}
\qquad &
\varphi^{(n)}_{\ell}(t)&\coloneqq
\varphi^{(n)}_{\ell\Re}( t)+\ii\, \varphi^{(n)}_{\ell\Im}(t),\ \  \ell=1,...,p, 
\end{align*}
%for the characteristic functions and their empirical counterparts.  
where the subscript $\Re$ (resp. $\Im$) is used for the real part (resp. imaginary part) of CFs and empirical CFs. Then notice that, under ${\cal H}_0$, the real part $\varphi^{(n)}_{\Re}$ of $\varphi^{(n)}$ admits the %two--term
expansion 
\begin{eqnarray}\nonumber    
\sqrt{n} \varphi^{(n)}_{\Re}(\boldsymbol t)&=&
\frac{1}{\sqrt{n}} \sum_{j=1}^n \cos(\boldsymbol {\widehat u}^\top_{n} \boldsymbol X_j) \\ \label{cosexp}&=&\frac{1}{\sqrt{n}} \sum_{j=1}^n \cos(\boldsymbol u^\top_0 \boldsymbol X_j) -\frac{1}{\sqrt{n}} \sum_{j=1}^n \sin(\boldsymbol u^\top_{0} \boldsymbol X_j) \boldsymbol X^\top_j (\boldsymbol {\widehat u}_{n}-\boldsymbol u_0)+{\rm{o}}_{\mathbb P}(1),
\end{eqnarray} 
where $\boldsymbol {\widehat u}_{n}=(\boldsymbol {\widehat \Omega}^{-1}_n)^\top \boldsymbol t$ %{I do not see any $\boldsymbol {\widehat u}_{n}$ in (\ref{conv})!},% 
and  $\boldsymbol u_0=(\boldsymbol \Omega^{-1}_0)^\top \boldsymbol t$, with $\boldsymbol \Omega_0$   denoting the actual mixing matrix under   the null hypothesis. Hereafter we assume $\bo {\Omega}_0=\bo {I}_p$, where ${\bo I}_p$ stands for the identity matrix of order $(p\times p)$. This does not imply any loss of generality in light of the affine-invariance property of Section \ref{invar}. As argued in Section \ref{invar}, permutation and sign matrices will likewise be suppressed below.

In this connection, assume that  
$\boldsymbol {\widehat \Omega}^{-1}_n$ 
%of the mixing matrix  $\boldsymbol \Omega$ satisfy
admits the Bahadur asymptotic representation
\begin{equation} \label{bahadur}
\sqrt{n} \left (\boldsymbol {\widehat \Omega}^{-1}_n-\boldsymbol I_p\right)^\top=\frac{1}{\sqrt{n}} \sum_{j=1}^n \boldsymbol V_{j}+{\rm{o}}_{\mathbb P}(1),
\end{equation}  
where the columns $\boldsymbol v_{j\ell}=\boldsymbol v_\ell(\boldsymbol Z_j)$ of 
 $(\boldsymbol V_{j}=\boldsymbol V(\boldsymbol Z_j), \ j=1,...,n)$ are   such that $\mathsf E[\boldsymbol v_{\ell}]=\boldsymbol 0$ and~$\mathsf E[\boldsymbol v_{\ell}\boldsymbol v^\top_{\ell}]<\infty$, $\ell=1,...,p$. Such expansions may be found for the  CF--based estimators of  \citet{ChenBickel2005}, as well as the R-estimators proposed by  \citet{IlmonenPaindaveine2011} and \citet{HallinMehta2015}.

Plugging \eqref{bahadur} into \eqref{cosexp}, and using the consistency of  empirical CFs, we have that   \begin{eqnarray} \label{cnexp}  
\sqrt{n} \left(\varphi^{(n)}_{\Re}(\boldsymbol t)-\varphi_\Re(\boldsymbol t)\right)&=&\frac{1}{\sqrt{n}} \sum_{j=1}^n \left(\cos(\boldsymbol t^\top  \boldsymbol Z_j)-\varphi_\Re(\boldsymbol t)\right) \\ \nonumber &+&  \frac{\partial \varphi_\Re(\boldsymbol t)}{\partial \boldsymbol t^\top}  \left( \frac{1}{\sqrt{n}}\sum_{j=1}^n \boldsymbol V_{j}\right)\boldsymbol t+{\rm{o}}_{\mathbb P}(1)\quad\text{as $n\to\infty$.}\end{eqnarray}

An analogous expansion also holds for the imaginary part $\varphi^{(n)}_{\Im}$ of $\varphi^{(n)}(\cdot)$, and this completes the asymptotic treatment of the first term in the right--hand side of \eqref{A.1}.

For the second term in the the right--hand side of the  decomposition \eqref{A.1} of  $\delta^{(n)}$,   first note that  \begin{eqnarray}   \nonumber && \sum_{\ell=1}^p \sqrt{n} (\varphi^{(n)}_{\ell}(t_\ell)-\varphi_{\ell}(t_\ell))\prod_{m<\ell} 
\varphi_{m}(t_m)\prod_{m>\ell} 
\varphi_{nm}(t_m)\\  \nonumber \ \ \ 
&&\qquad\qquad\qquad =\sum_{\ell=1}^p \sqrt{n} (\varphi^{(n)}_{\ell}(t_\ell)-\varphi_{\ell}(t_\ell))\prod_{m\neq \ell} 
\varphi_{m}(t_m)+ {\rm{o}}_{\mathbb P}(1) 
\\ \label{prodexp}  && \qquad\qquad\qquad =\sum_{\ell=1}^p \sqrt{n} (\varphi^{(n)}_{\ell}(t_\ell)-\varphi_{\ell}(t_\ell))\:\varphi_{(\ell)}(\boldsymbol t_{(\ell)})+ {\rm{o}}_{\mathbb P}(1),
\end{eqnarray}
where $\varphi_{(\ell)}(\boldsymbol t_{(\ell)})\coloneqq \varphi_{(\ell)\Re}(\boldsymbol t_{(\ell)})+\ii \varphi_{(\ell)\Im}(\boldsymbol t_{(\ell)})$ denotes the joint CF of~$(Z_m, \ m\neq \ell)$, computed at  $\boldsymbol t_{(\ell)}\coloneqq (t_1,...,t_{\ell-1},t_{\ell+1},...,t_p)^\top \in \mathbb R^{p-1}$, $\ell=1,...,p$.

We now treat the real part  of the $\ell^{\rm{th}}$ summand    $\varphi^{(n)}_{\ell\Re}$ in \eqref{prodexp}. Specifically,   by taking a two-term expansion and using similar arguments as in \eqref{cosexp}, we have 
\begin{eqnarray} \label{prodexp1}   
\sqrt{n} \left(\varphi^{(n)}_{\ell\Re}(t_\ell)-\varphi_{\ell\Re}(t_\ell)\right)&=&\frac{1}{\sqrt{n}} \sum_{j=1}^n \left(\cos(t_{\ell } Z_{j\ell})-\varphi_{\ell\Re}(t_\ell)\right) \\ \nonumber 
&+&  
\frac{\partial \varphi_\Re(\boldsymbol t_\ell)}{\partial \boldsymbol t^\top} \left( \frac{1}{\sqrt{n}}\sum_{j=1}^n \boldsymbol v_{j\ell}\right) t_\ell+{\rm{o}}_{\mathbb P}(1),
\end{eqnarray}
where $\boldsymbol t_\ell\coloneqq (0,...,0,t_\ell,0,...,0)^\top$, $\ell=1,...,p$.  
Using   \eqref{cosexp}, \eqref{cnexp}  and \eqref{prodexp1} and   analogous expansions for the imaginary parts, we obtain    $\delta^{(n)}_{m} = Z^{(n)*}_{m}+{\rm{o}}_{\mathbb  P}(1)$ with $m=\Re$ or~$\Im$,  where 
%\begin{eqnarray*}\nonumber   
\[
Z^{(n)*}_{m}(\boldsymbol t)=\frac{1}{\sqrt{n}} \sum_{j=1}^n Y_{jm}(\boldsymbol t)\quad  
%\end{eqnarray*}
\text{and}\quad Y_{jm}(\boldsymbol t)\coloneqq Y_m(\boldsymbol Z_j;\boldsymbol t)
\]
with
\begin{eqnarray*}\nonumber 
Y_{\Re}(\boldsymbol Z;\boldsymbol t)&=& \left(\cos(\boldsymbol t^\top \boldsymbol Z)-\varphi_\Re(\boldsymbol t)\right)+\frac{\partial \varphi_\Re(\boldsymbol t)}{\partial \boldsymbol t^\top}\:   \boldsymbol V(\boldsymbol Z) \boldsymbol t \\ 
&+&\sum_{\ell=1}^p \left\{ \left(\sin(t_{\ell } Z_{\ell})-\varphi_{\ell\Im}(t_\ell)\right)  + \frac{\partial \varphi_\Im(\boldsymbol t_\ell)}{\partial \boldsymbol t^\top}\:   \boldsymbol v_{\ell}(\boldsymbol Z) t_\ell \right \} \varphi_{(\ell)\Im}(\boldsymbol t_{(\ell)}) \\
&-& \left\{ \left(\cos(t_{\ell}Z_\ell)-\varphi_{\ell\Re}(t_\ell)\right)
+\frac{\partial \varphi_\Re(\boldsymbol t_\ell)}{\partial \boldsymbol t^\top}\:  \boldsymbol v_{\ell}(\boldsymbol Z) t_\ell \right \} \varphi_{(\ell)\Re}(\boldsymbol t_{(\ell)})
\end{eqnarray*}
and
\begin{eqnarray*}\nonumber 
Y_{\Im}(\boldsymbol Z;\boldsymbol t)&=& \left(\sin(\boldsymbol t^\top \boldsymbol Z)-\varphi_\Im(\boldsymbol t)\right)+ \frac{\partial \varphi_\Im(\boldsymbol t)}{\partial \boldsymbol t^\top}\:  \boldsymbol V(\boldsymbol Z) \boldsymbol t \\ 
&-&\sum_{\ell=1}^p \left\{ \left(\cos(t_{\ell } Z_{\ell})-\varphi_{\ell\Re}(t_\ell)\right) 
\frac{\partial \varphi_\Re(\boldsymbol t_\ell)}{\partial \boldsymbol t^\top}\:   \boldsymbol v_{\ell}(\boldsymbol Z) t_\ell \right \} \varphi_{(\ell)\Im}(\boldsymbol t_{(\ell)}) \\
&+& \left\{ \left(\sin(t_{\ell}Z_\ell)-\varphi_{\ell\Im}(t_\ell)\right)
+\frac{\partial \varphi_\Im(\boldsymbol t_\ell)}{\partial \boldsymbol t^\top}\:  \boldsymbol v_{\ell}(\boldsymbol Z) t_\ell \right \} \varphi_{(\ell)\Re}(\boldsymbol t_{(\ell)}).
\end{eqnarray*}

Hence, under the null hypothesis ${\cal{H}}_0$, we have, by the Central Limit Theorem (see, e.g., Theorem 2.7 in  \citet{Bosq2000}), that~$\delta^{(n)}_{m}(\boldsymbol t)\verk~\!\delta_m(\boldsymbol t)$,  $m=\Re$ or $\Im$, pointwise\linebreak in~$\boldsymbol t\in \mathbb R^p$, where $\{\delta_m\}$ is a zero--mean Gaussian process with covariance kernel 
$$K_m(\boldsymbol s,\boldsymbol t)=\mathsf E[Y_m(\boldsymbol s)Y_m(\boldsymbol t)], \quad m=\Re\  \text{or} \ \Im,
$$ and consequently $$\delta^{(n)}(\boldsymbol t)\verk \delta(\boldsymbol t):=\delta_\Re(\boldsymbol t)+\imath \delta_\Im(\boldsymbol t).$$

\begin{rem}
Clearly, each  of the aforementioned covariance kernels $K_{\Re}(\boldsymbol s,\boldsymbol t)$\linebreak and $K_{\Im}(\boldsymbol s,\boldsymbol t)$   depends on the (joint) characteristic function $\varphi$ of $\boldsymbol Z$, as well as  the corresponding marginal CFs $(\varphi_\ell, \ \ell=1,...,p)$. Moreover, the type of estimator used for $\bf \Omega$  also enters these kernels via the Bahadur expansion~\eqref{bahadur}. These facts make the limiting null distribution intractable for actual test implementation. Even if the joint and marginal characteristic functions in $\bo Z$ were known, the asymptotic null distribution of the test statistic~$T_{n,W}$  would still remain highly non--standard---the distribution of a weighted infinite sum of  i.i.d.~chi--squared random variables with one degree of freedom, with weights requiring the computation of the eigenvalues of a certain integral equation; see for instance \citet{PfisterBuhlmannScholkopfPeters2017}, Section~3.1.          
\end{rem}

%\newpage

\section{Additional Simulation results}  \label{AppB}

In this appendix, we provide the simulation results for Setting~1 that could not be included in Section~\ref{sec_7} of the main text.

\subsection{Additional results for Setting 1}  \label{AppB1}

\begin{table}[ht!]
\centering
\footnotesize
\begin{tabular*}{\textwidth}{@{\extracolsep{\fill}}*{9}{c}}
  \hline
  &  \multicolumn{4}{c}{$T_{n,L}^{\text{\rm Boot}}$} &  \multicolumn{4}{c}{$T_{n,L}^{\text{\rm Perm}}$}  \\ 
   \cline{2-5}\cline{6-9}
n & JADE & FOBI & FastICA & TRUE & JADE & FOBI & FastICA & TRUE\\ 
  \hline
500 & 0.0530 & 0.1170 & 0.0471 & 0.0550 & 0.0780 & 0.0980 & 0.0481 & 0.0660 \\ 
  1,000 & 0.0730 & 0.1090 & 0.0520 & 0.0540 & 0.0420 & 0.1160 & 0.0450 & 0.0540 \\ 
  2,000 & 0.0610 & 0.1060 & 0.0320 & 0.0500 & 0.0520 & 0.1020 & 0.0460 & 0.0650 \\ 
  4,000 & 0.0610 & 0.0840 & 0.0490 & 0.0660 & 0.0530 & 0.0930 & 0.0480 & 0.0530 \\ 
  8,000 & 0.0420 & 0.0530 & 0.0690 & 0.0480 & 0.0840 & 0.0590 & 0.0490 & 0.0560 \\ 
  16,000 & 0.0560 & 0.0230 & 0.0440 & 0.0410 & 0.0610 & 0.0200 & 0.0380 & 0.0440 \\ 
   \hline
\end{tabular*}
\caption{Empirical sizes (rejection frequencies over 1,000 replications) of the tests based on  $T_{n,L}^{\text{\rm Boot}}$ and $T_{n,L}^{\text{\rm Perm}}$
%%, $T_{n,\text{DC}}$, $\tenq{T}_{n,Id,G}^{\text{\rm Boot}}$, and~$\tenq{T}_{n,Id, G}^{\text{\rm Perm}}$
  (nominal size $\alpha = 5\%$) under Setting 1 and 
%%for  $T_{G,B}, T_{G,P}$ and $T_{DC}$ and for
  various ICA methods. Column TRUE corresponds to the oracle test based on the actual values of~the~$\bo Z$'s.
} \label{PowerH0lap}
\end{table}

\begin{table}[h!]
\centering
\footnotesize
\begin{tabular*}{\textwidth}{@{\extracolsep{\fill}}*{10}{c}}
  \hline
  &  \multicolumn{4}{c}{$\tenq{T}_{n,\text{Id},L}^{\text{\rm Boot}}$} &  \multicolumn{4}{c}{$\tenq{T}_{n,\text{Id}, L}^{\text{\rm Perm}}$}  \\ 
   \cline{2-5}\cline{6-9}
n & JADE & FOBI & FastICA & TRUE & JADE & FOBI & FastICA & TRUE\\ 
  \hline
500 & 0.1090 & 0.1470 & 0.0603 & 0.0720 & 0.0870 & 0.1470 & 0.0672 & 0.0550 \\ 
  1,000 & 0.0600 & 0.1480 & 0.0410 & 0.0470 & 0.0680 & 0.1590 & 0.0380 & 0.0420 \\ 
  2,000 & 0.0520 & 0.1260 & 0.0430 & 0.0590 & 0.0750 & 0.1200 & 0.0480 & 0.0430 \\ 
  4,000 & 0.0710 & 0.0880 & 0.0410 & 0.0450 & 0.0580 & 0.1070 & 0.0440 & 0.0570 \\ 
  8000 & 0.0860 & 0.0650 & 0.0540 & 0.0410 & 0.0500 & 0.0680 & 0.0440 & 0.0310 \\ 
  16,000 & 0.0440 & 0.0220 & 0.0670 & 0.0470 & 0.0550 & 0.0290 & 0.0510 & 0.0600 \\ 
   \hline
\end{tabular*}
\caption{Empirical sizes (rejection frequencies over 1,000 replications) of the tests based on   $\tenq{T}_{n,\text{Id},L}^{\text{\rm Boot}}$ and~$\tenq{T}_{n,\text{Id}, L}^{\text{\rm Perm}}$  (nominal size $\alpha = 5\%$) under Setting 1 and 
%for  $T_{G,B}, T_{G,P}$ and $T_{DC}$ and for
  various ICA methods. Column TRUE corresponds to the oracle test based on   the actual  values of~the~$\bo Z$'s.} \label{PowerH0rl}
\end{table}

$\,$\vspace{6mm}

\begin{table}[h!]
\centering
\footnotesize
\begin{tabular*}{\textwidth}{@{\extracolsep{\fill}}*{9}{c}}
  \hline
  &  \multicolumn{4}{c}{$\tenq{T}_{n,\text{vdW},G}^{\text{\rm Boot}}$} &  \multicolumn{4}{c}{$\tenq{T}_{n,\text{vdW},G}^{\text{\rm Perm}}$}  \\
   \cline{2-5}\cline{6-9}
n & JADE & FOBI & FastICA & TRUE & JADE & FOBI & FastICA & TRUE\\ 
  \hline
500 & 0.0970 & 0.1830 & 0.0724 & 0.0670 & 0.0900 & 0.1650 & 0.0653 & 0.0730 \\ 
  1,000 & 0.0640 & 0.1510 & 0.0500 & 0.0600 & 0.0630 & 0.1620 & 0.0520 & 0.0570 \\ 
  2,000 & 0.0570 & 0.1100 & 0.0420 & 0.0630 & 0.0710 & 0.1150 & 0.0500 & 0.0660 \\ 
  4,000 & 0.0770 & 0.1130 & 0.0460 & 0.0550 & 0.0630 & 0.0870 & 0.0420 & 0.0610 \\ 
  8,000 & 0.0510 & 0.0520 & 0.0570 & 0.0310 & 0.0540 & 0.0580 & 0.0620 & 0.0410 \\ 
  16,000 & 0.0560 & 0.0300 & 0.0660 & 0.0610 & 0.0660 & 0.0300 & 0.0470 & 0.0630 \\ 
   \hline
\end{tabular*}
\caption{Empirical sizes (rejection frequencies over 1,000 replications) of the tests based  on~$\tenq{T}_{n,\text{vdW},G}^{\text{\rm Boot}}$ and~$\tenq{T}_{n,\text{vdW},G}^{\text{\rm Perm}}$  (nominal size $\alpha = 5\%$) under Setting 1 and 
%for  $T_{G,B}, T_{G,P}$ and $T_{DC}$ and for
  various ICA methods. Column TRUE corresponds to the oracle test based on   the actual  values of the $\bo Z$'s.} \label{PowerH0vdW}
\end{table}
%\clearpage
\newpage

\subsection{Additional results for Setting 2}  \label{AppB2}

\begin{figure}[h!]
    \centering
    \includegraphics[width=0.73\textwidth]{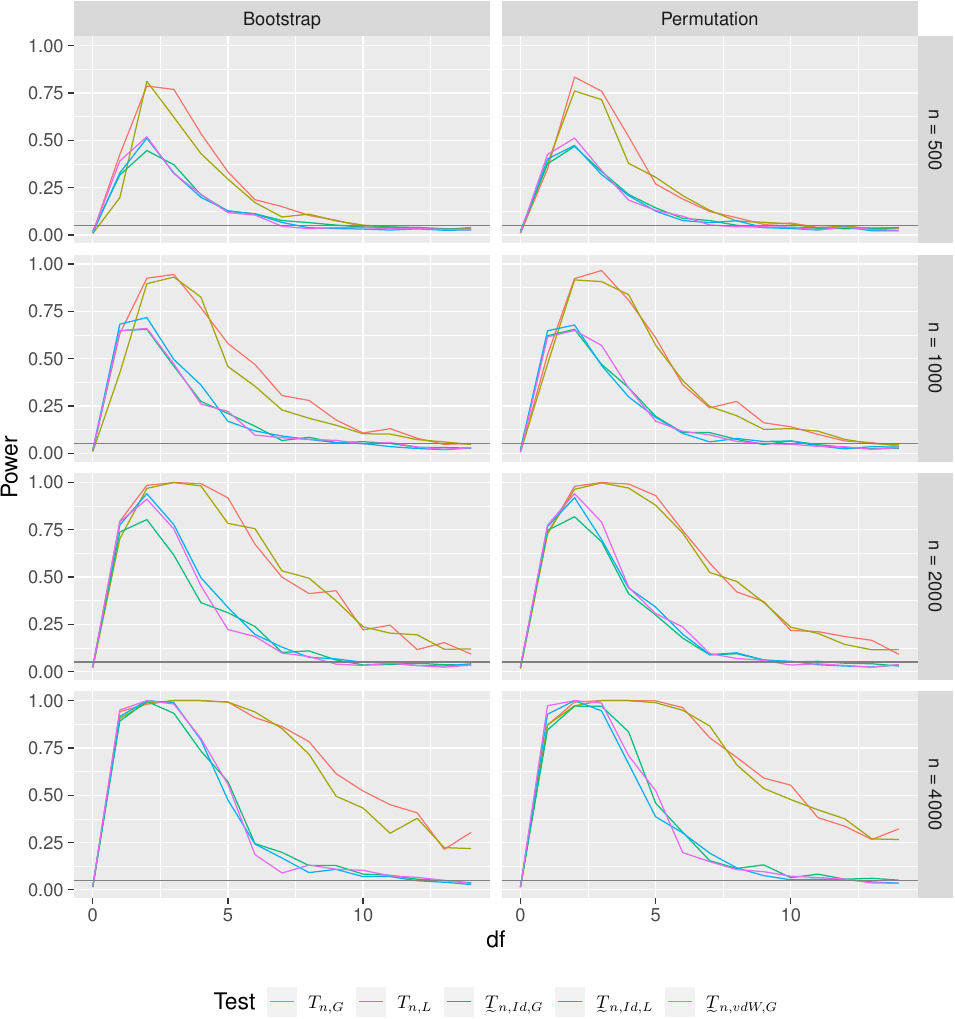}
    \caption{\small Empirical powers (rejection frequencies over 1,000 replications) of the tests based on~$T_{n,G}^{\text{\rm Boot}}$,  $T_{n,L}^{\text{\rm Boot}}$,  $\tenq{T}_{n,\text{Id},G}^{\text{\rm Boot}}$,   $\tenq{T}_{n,\text{Id},L}^{\text{\rm Boot}}$, $\tenq{T}_{n,\text{vdW},G}^{\text{\rm Boot}}$, $T_{n,G}^{\text{\rm Perm}}$,  $T_{n,L}^{\text{\rm  Perm}}$,  $\tenq{T}_{n,\text{Id},G}^{\text{\rm  Perm}}$,   $\tenq{T}_{n,\text{Id},L}^{\text{\rm  Perm}}$, and $\tenq{T}_{n,\text{vdW},G}^{\text{\rm Perm}}$,  
       for sample sizes $n=500,\, 1,000,\, 2,000$, and $4,000$ in Setting 2 (spherical $t$), as functions of the degrees of freedom when using FOBI, with bootstrap and permutational critical values, respectively. The grey horizontal line represents the nominal size $\alpha = 5\%$. Note that in the figure the Gaussian distribution ($t_\infty$) is represented under df=0.}
    \label{fig:PowerS2Fobi}
\end{figure}

\begin{figure}[ht]
    \centering
    \includegraphics[width=0.73\textwidth]{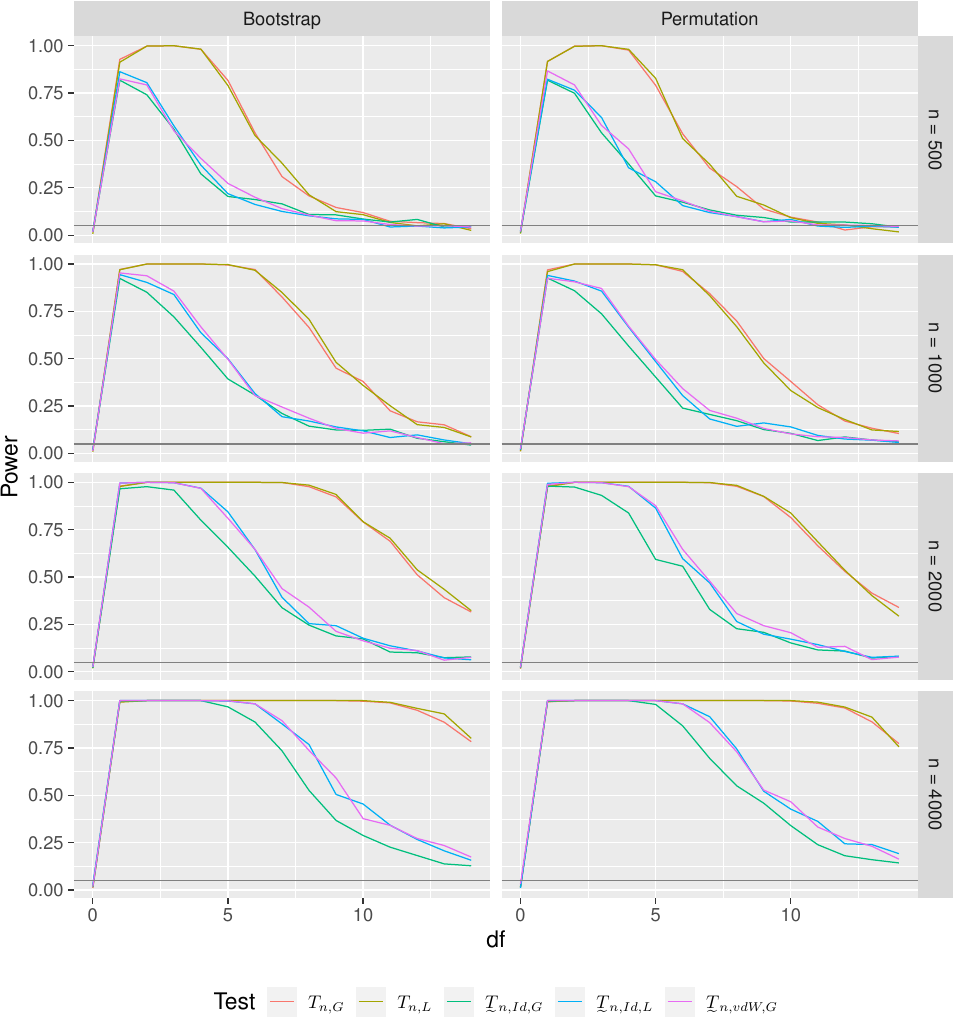}
    \caption{\small Empirical powers (rejection frequencies over 1,000 replications) of the tests based on~$T_{n,G}^{\text{\rm Boot}}$,  $T_{n,L}^{\text{\rm Boot}}$,  $\tenq{T}_{n,\text{Id},G}^{\text{\rm Boot}}$,   $\tenq{T}_{n,\text{Id},L}^{\text{\rm Boot}}$, $\tenq{T}_{n,\text{vdW},G}^{\text{\rm Boot}}$, $T_{n,G}^{\text{\rm Perm}}$,  $T_{n,L}^{\text{\rm  Perm}}$,  $\tenq{T}_{n,\text{Id},G}^{\text{\rm  Perm}}$,   $\tenq{T}_{n,\text{Id},L}^{\text{\rm  Perm}}$, and $\tenq{T}_{n,\text{vdW},G}^{\text{\rm Perm}}$ 
       for sample sizes $n=500,\, 1,000,\, 2,000$, and $4,000$ in Setting 2 (spherical $t$), as functions of the degrees of freedom when using JADE, with bootstrap and permutational critical values, respectively. The grey horizontal line represents the nominal size $\alpha = 5\%$. Note that in the figure the Gaussian distribution ($t_\infty$) is represented under df=0.}
    \label{fig:PowerS2Jade}
\end{figure}

$\,$

\clearpage

\newpage

\subsection{Additional results for Setting 3}  \label{AppB3}
\begin{figure}[h!]
    \centering
    \includegraphics[width=0.73\textwidth]{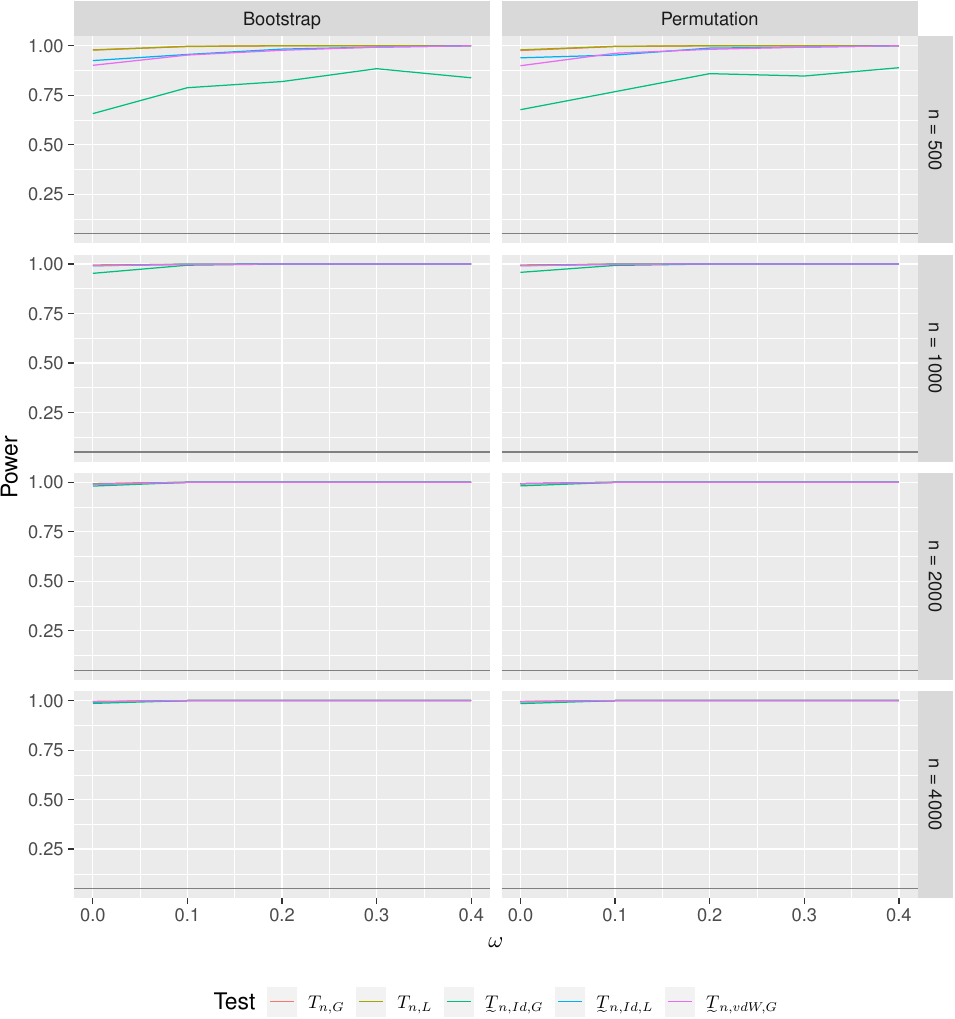}
    \caption{\small Empirical powers (rejection frequencies over 1,000 replications) of the tests based on~$T_{n,G}^{\text{\rm Boot}}$,  $T_{n,L}^{\text{\rm Boot}}$,  $\tenq{T}_{n,\text{Id},G}^{\text{\rm Boot}}$,   $\tenq{T}_{n,\text{Id},L}^{\text{\rm Boot}}$, $\tenq{T}_{n,\text{vdW},G}^{\text{\rm Boot}}$, $T_{n,G}^{\text{\rm Perm}}$,  $T_{n,L}^{\text{\rm  Perm}}$,  $\tenq{T}_{n,\text{Id},G}^{\text{\rm  Perm}}$,   $\tenq{T}_{n,\text{Id},L}^{\text{\rm  Perm}}$, and $\tenq{T}_{n,\text{vdW},G}^{\text{\rm Perm}}$
    %, and~$T_{n,\text{DC}}$
       for sample  sizes $n=500,\, 1,000,\, 2,000$, and $4,000$ in Setting 3 (Clayton copula), as  functions of the copula  parameter $\omega$  when using FOBI and bootstrap or permutational critical values, respectively. The grey horizontal line represents the nominal size $\alpha = 5\%$.}
    \label{fig:PowerS3Fobi}
\end{figure}
\newpage

\begin{figure}[h!]
    \centering
    \includegraphics[width=0.73\textwidth]{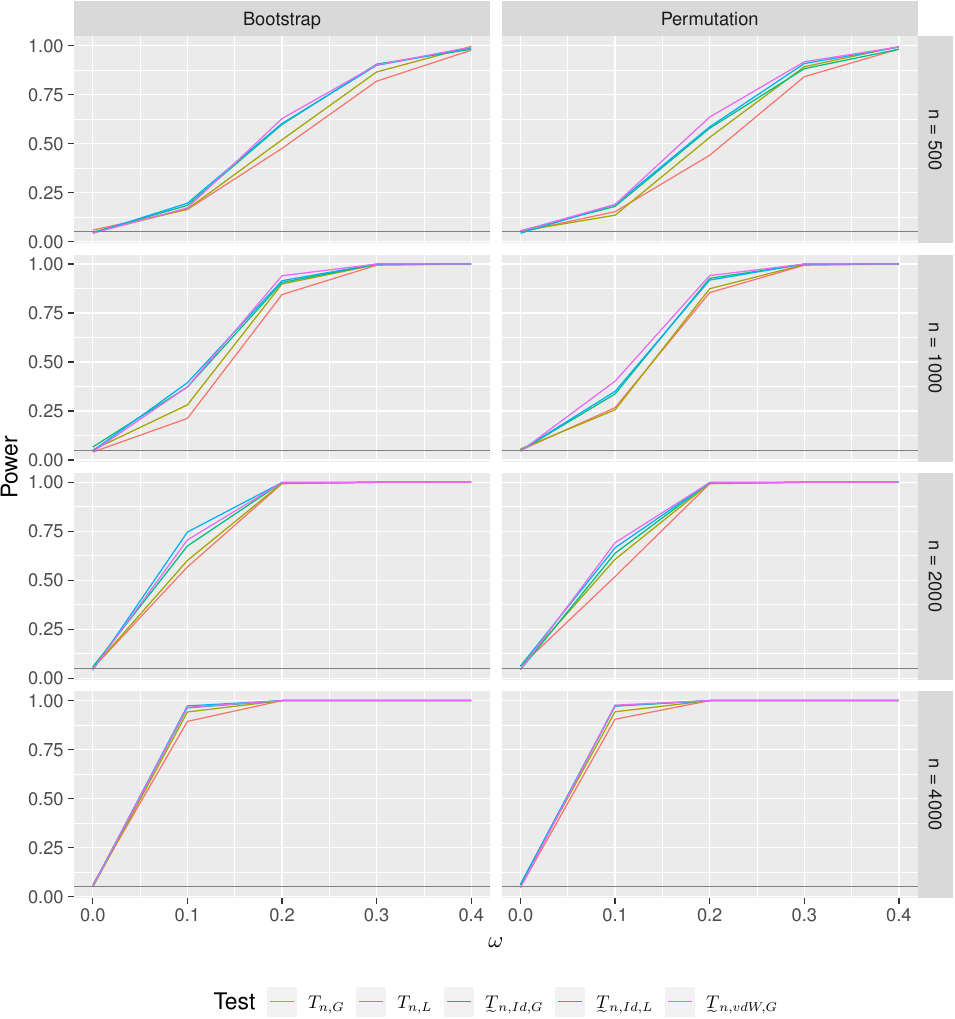}
    \caption{\small Empirical powers (rejection frequencies over 1,000 replications) of the tests based on~$T_{n,G}^{\text{\rm Boot}}$,  $T_{n,L}^{\text{\rm Boot}}$,  $\tenq{T}_{n,\text{Id},G}^{\text{\rm Boot}}$,   $\tenq{T}_{n,\text{Id},L}^{\text{\rm Boot}}$, $\tenq{T}_{n,\text{vdW},G}^{\text{\rm Boot}}$, $T_{n,G}^{\text{\rm Perm}}$,  $T_{n,L}^{\text{\rm  Perm}}$,  $\tenq{T}_{n,\text{Id},G}^{\text{\rm  Perm}}$,   $\tenq{T}_{n,\text{Id},L}^{\text{\rm  Perm}}$, and $\tenq{T}_{n,\text{vdW},G}^{\text{\rm Perm}}$
    %, and~$T_{n,\text{DC}}$
       for sample sizes $n=500,\, 1,000,\, 2,000$, and $4,000$ in Setting 3 (Clayton copula), as functions of the copula parameter $\omega$  when using JADE  and bootstrap or permutational critical values, respectively. The grey horizontal line represents the nominal size $\alpha = 5\%$.}
    \label{fig:PowerS3Jade}
\end{figure}

\clearpage

\subsection{Comments about {\it\bf warp-speed} methods}  \label{AppB4}

To make simulations feasible we used the {\it warp-speed bootstrap} approach of \citet{GiacominiPolitisWhite2013} in which in each simulation iteration only one bootstrap sample is created. Then the bootstrap samples from all $M$ iterations are combined and can be used to compute the critical value or the $p$-value  of interest. 
%, in our case the $p$-value of the test statistic. 
As in our simulation settings~$M=~\!1,000$, the simulation results in the previous sections are based on a   number~$M=1,000$ of bootstrap samples. To evaluate the impact of the number of bootstrap samples, we also checked how in Setting~1 the size changes when using only a random sample out of these $M=1,000$ warp-speed bootstrap samples. The results are shown in Tables~\ref{PowerH0gauss1000}-\ref{PowerH0gauss200} and reveal that 200 samples might not be sufficient for the approximation while 500 seems to be a good value. For other score functions, the results, which are not shown here,  are quite comparable.

\begin{table}[ht]
\centering
\footnotesize
\begin{tabular*}{\textwidth}{@{\extracolsep{\fill}}*{9}{c}}
  \hline
  &  \multicolumn{4}{c}{
  $\tenq{T}^{\text{\rm Boot}}_{n, \text{Id},G}$
 }
 & 
   \multicolumn{4}{c}{
   $\tenq{T}^{\text{\rm Perm}}_{n, \text{Id},G
   }$ 
   }
   \\ 
   \cline{2-5}\cline{6-9}
$n$ & JADE & FOBI & FastICA & TRUE & JADE & FOBI & FastICA & TRUE\\ 
  \hline
500 & 0.0700 & 0.1400 & 0.0602 & 0.0640 & 0.0590 & 0.1310 & 0.0453 & 0.0580 \\ 
  1,000 & 0.0540 & 0.1350 & 0.0400 & 0.0640 & 0.0690 & 0.1260 & 0.0500 & 0.0470 \\ 
  2,000 & 0.0600 & 0.1040 & 0.0590 & 0.0750 & 0.0560 & 0.1180 & 0.0510 & 0.0750 \\ 
  4,000 & 0.0630 & 0.0910 & 0.0520 & 0.0570 & 0.0730 & 0.0840 & 0.0430 & 0.0530 \\ 
  8,000 & 0.0700 & 0.0650 & 0.0660 & 0.0680 & 0.0700 & 0.0540 & 0.0610 & 0.0380 \\ 
  16,000 & 0.0500 & 0.0390 & 0.0420 & 0.0550 & 0.0500 & 0.0300 & 0.0480 & 0.0670 \\ 
   \hline
\end{tabular*}
\caption{Empirical sizes (rejection frequencies over 1,000 replications) of the tests based on   $\tenq{T}^{\text{\rm Boot}}_{n, \text{Id},G}$ and~$\tenq{T}^{\text{\rm Perm}}_{n, \text{Id},G
   }$  (nominal size $\alpha = 5\%$) under Setting 1 and 
%for  $T_{G,B}, T_{G,P}$ and $T_{DC}$ and for
  various ICA methods. Column TRUE corresponds to the oracle test based on the actual values of the $\bo Z$'s.
 Critical values are based on 1,000 bootstrap samples and 1,000 permutations, respectively.} \label{PowerH0gauss1000}
\end{table}

\begin{table}[ht]
\centering
\footnotesize,
\begin{tabular*}{\textwidth}{@{\extracolsep{\fill}}*{9}{c}}
  \hline
  &  \multicolumn{4}{c}{$T_{rG,B}$} &  \multicolumn{4}{c}{$T_{rG,P}$}  \\ 
   \cline{2-5}\cline{6-9}
n & JADE & FOBI & FastICA & TRUE & Jade & Fobi & FastICA & TRUE\\ 
  \hline
500 & 0.0700 & 0.1560 & 0.0562 & 0.0750 & 0.0760 & 0.1310 & 0.0453 & 0.0510 \\ 
  1000 & 0.0550 & 0.1400 & 0.0320 & 0.0560 & 0.0600 & 0.1220 & 0.0430 & 0.0460 \\ 
  2000 & 0.0560 & 0.1120 & 0.0620 & 0.0840 & 0.0490 & 0.1260 & 0.0460 & 0.0780 \\ 
  4000 & 0.0490 & 0.0850 & 0.0600 & 0.0580 & 0.0730 & 0.0860 & 0.0380 & 0.0540 \\ 
  8000 & 0.0630 & 0.0710 & 0.0720 & 0.0680 & 0.0600 & 0.0590 & 0.0630 & 0.0350 \\ 
  16000 & 0.0480 & 0.0390 & 0.0570 & 0.0580 & 0.0370 & 0.0250 & 0.0510 & 0.0670 \\ 
   \hline
\end{tabular*}
\caption{Empirical sizes (rejection frequencies over 1,000 replications) of the tests based on   $\tenq{T}^{\text{\rm Boot}}_{n, \text{Id},G}$ and~$\tenq{T}^{\text{\rm Perm}}_{n, \text{Id},G
   }$  (nominal size $\alpha = 5\%$) under Setting 1 and 
%for  $T_{G,B}, T_{G,P}$ and $T_{DC}$ and for
  various ICA methods. Column TRUE corresponds to the oracle test based on the actual values of the $\bo Z$'s.
 Critical values are based on 500 bootstrap samples and 500 permutations, respectively.\vspace{2mm}} \label{PowerH0gauss500}
\end{table}

\begin{table}[ht]
\centering
\footnotesize
\begin{tabular*}{\textwidth}{@{\extracolsep{\fill}}*{9}{c}}
  \hline
  &  \multicolumn{4}{c}{$\tenq{T}^{\text{\rm Boot}}_{n, \text{Id},G}$ } &  \multicolumn{4}{c}{$\tenq{T}^{\text{\rm Perm}}_{n, \text{Id},G
   }$}  \\ 
  \cline{2-5}\cline{6-9}
n & JADE & FOBI & FastICA & TRUE & Jade & Fobi & FastICA & TRUE\\ 
  \hline
500 & 0.0970 & 0.1490 & 0.0602 & 0.0660 & 0.0760 & 0.0840 & 0.0735 & 0.0580 \\ 
  1000 & 0.0310 & 0.0920 & 0.0420 & 0.0610 & 0.0640 & 0.1240 & 0.0600 & 0.0410 \\ 
  2000 & 0.0420 & 0.1100 & 0.0560 & 0.0750 & 0.0740 & 0.1500 & 0.0630 & 0.0750 \\ 
  4000 & 0.0260 & 0.0970 & 0.0690 & 0.0160 & 0.0830 & 0.0920 & 0.0320 & 0.0580 \\ 
  8000 & 0.0700 & 0.0600 & 0.0720 & 0.0530 & 0.0540 & 0.0540 & 0.0630 & 0.0540 \\ 
  16000 & 0.0540 & 0.0190 & 0.0310 & 0.0560 & 0.0500 & 0.0190 & 0.0380 & 0.0670 \\ 
   \hline
\end{tabular*}
\caption{Empirical sizes (rejection frequencies over 1,000 replications) of the tests based on   $\tenq{T}^{\text{\rm Boot}}_{n, \text{Id},G}$ and~$\tenq{T}^{\text{\rm Perm}}_{n, \text{Id},G
   }$  (nominal size $\alpha = 5\%$) under Setting 1 and 
%for  $T_{G,B}, T_{G,P}$ and $T_{DC}$ and for
  various ICA methods. Column TRUE corresponds to the oracle test based on the actual values of the $\bo Z$'s. 
 Critical values are based on 200 bootstrap samples and 200 permutations, respectively.\vspace{5mm}} \label{PowerH0gauss200}
\end{table}

\begin{table}[h!]
\centering
\footnotesize
\begin{tabular*}{\textwidth}{@{\extracolsep{\fill}}*{7}{c}}
$P$ & Power $T^{\text{\rm wsPerm}}_{n,G}$ & Power $T^{\text{\rm fPerm}}_{n,G}$ & Disagree & Power $\tenq{T}^{\text{\rm wsPerm}}_{n,\text{Id},G}$ & Power $\tenq{T}^{\text{\rm fPerm}}_{n,\text{Id},G}$ & Disagree \\
\hline
1,000            & 0.054             & 0.057            & 1.5 \% & 0.068              & 0.072              & 2.4\%  \\
500             & 0.053             & 0.053            & 2.0 \% & 0.072              & 0.072              & 2.4\%  \\
200             & 0.057             & 0.052            & 2.3 \% & 0.069              & 0.059              & 2.8\% \\
\hline
\end{tabular*}
\caption{Comparison of warp-speed (wsPerm) permutation tests versus full (fPerm) permutation tests for different numbers of permutations $P$ used in Setting~1 with~$n=2,000$ based on 1,000 replications. The column ``Disagree'' gives the percentage of cases where the test decisions in the two variants differed at the level $\alpha=0.05$.} \label{tab:CompWsPfP}\vspace{5mm}
\end{table}

While warp-speed bootstrap has a sound theoretical  foundation \citep{GiacominiPolitisWhite2013}, it seems that warp-speed permutations have not been considered so far in the literature. We leave the theoretic justification for further research and only show below some empirical evidence that supports such an approach. In all simulations shown above, the warp-speed permutation $p$--values are computed such that, at each itera\-tion~$m=1,\ldots,M$, only one permutation is performed, and all these permutations are combined for the individual $p$--values. While in the traditional, full permutational, approach at each iteration $P$ permutation should be executed, and the $p$--value at that iteration based on these permutations only. As in our simulations study $M=1,000$ above all warp-speed permutations are therefore based on a comparison to 1000 reference values. Assuming that like in the bootstrap case, as investigated in Tables~\ref{PowerH0gauss1000}-\ref{PowerH0gauss200}, also a smaller number $P$ of permutations might be of interest, we considered smaller values for~$P$ and restricted   the warp-speed $p$--value computation to a random subsample of the $M$ permuation values of size $P$. We did this comparison in Setting 1 using the tests
$T_{n,G}$ and $\tenq{T}_{n,\text{Id},G}$ using  $n=2,000$ based on~$M=1,000$ replications. The   FastICA method was used here. The results are presented in Table~\ref{tab:CompWsPfP}. The table reports also the percentage of cases where the full permutation case and the warp-speed case lead to different decisions at probability level $\alpha=0.05$. As the table shows, the differences in overall power between the full permutational approach and the warp-speed approach are minimal and lead in almost all cases to the same decision,  thereby fully justifying the usage of the warp-speed approach in our simulation study. The number of permutations seems to have little impact here but we nevertheless recommend to use rather more than 200.

%\newpage 

\section{Additional results for the  application in Section~8}  \label{AppC}

\begin{figure}[hb!]
    \centering
    \includegraphics[width=0.8\textwidth]{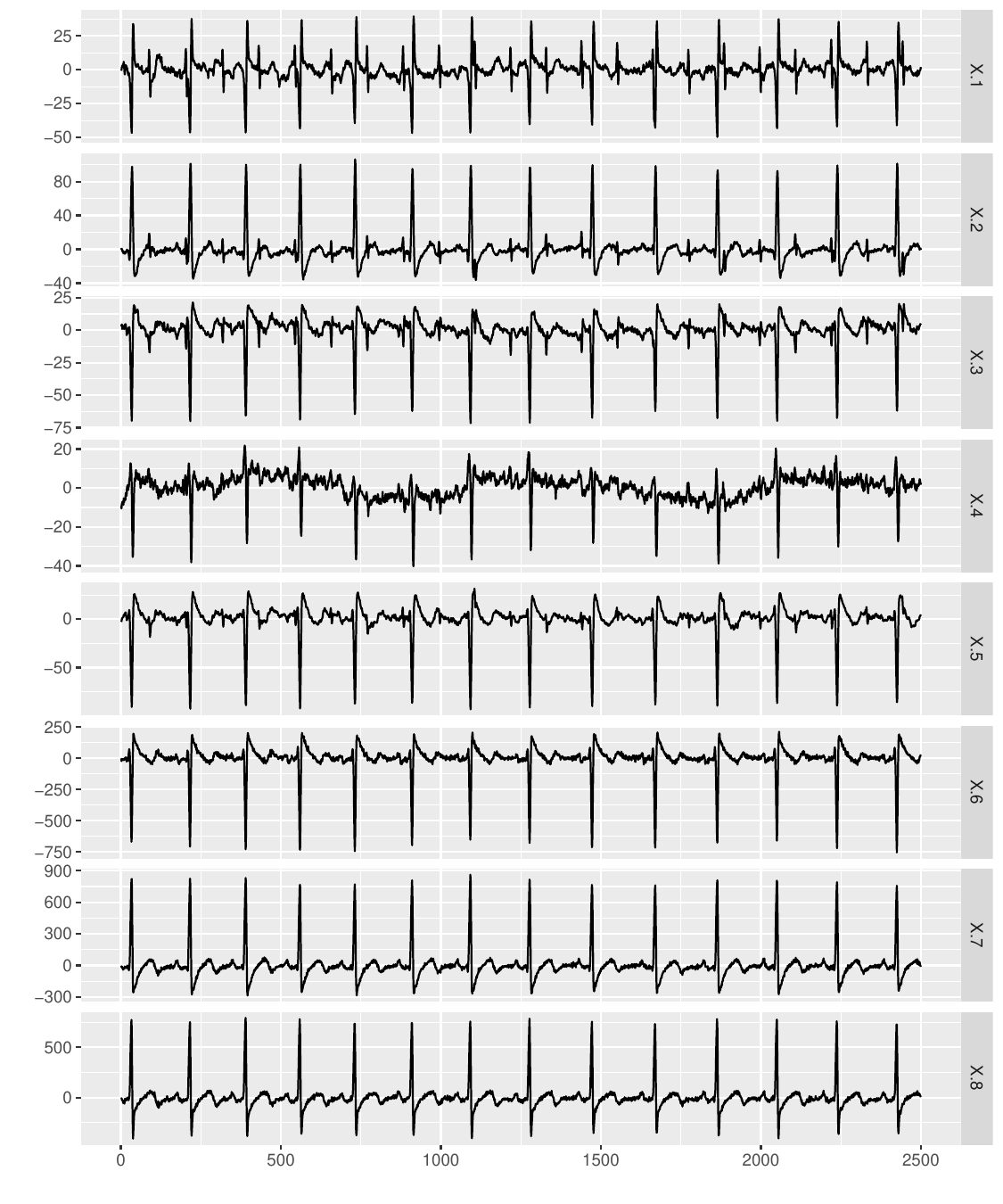}
    \caption{ECG measurements for the pregnant women.}
    \label{fig:ECGdata}
\end{figure}

\begin{figure}
    \centering
    \includegraphics[width=0.8\textwidth]{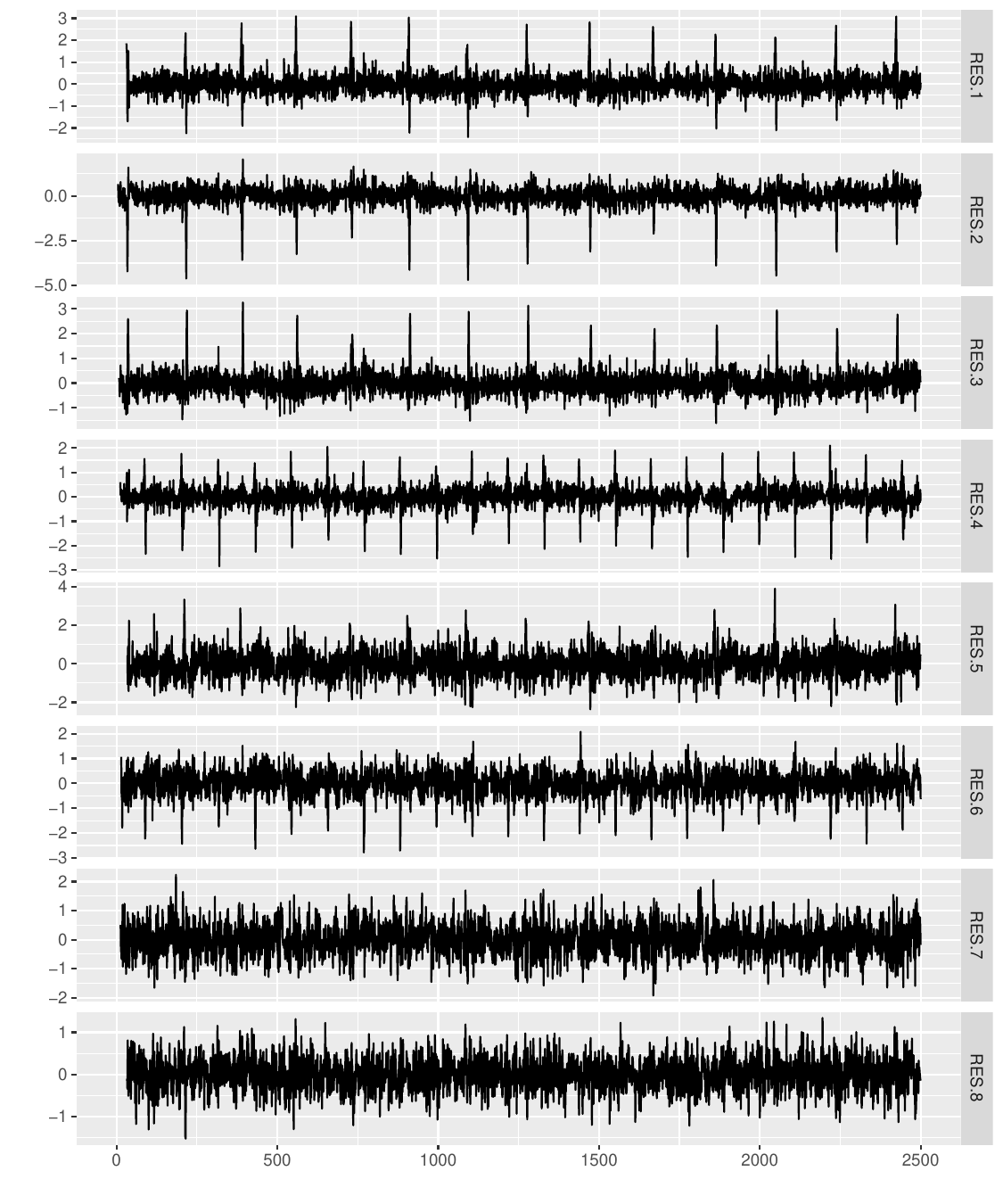}
    \caption{Residuals after fitting AR processes to all independent components based on JADE for the ECG measurements of the pregnant women.}
    \label{fig:ECGres}
\end{figure}

\clearpage

%\bibliographystyle{abbrvnat}
%\bibliography{Refs}%}

\end{document}